\documentclass{article}

\usepackage{hyperref}
\usepackage[sort&compress,numbers]{natbib}
\usepackage[
    type={CC},
    modifier={by},
    version={4.0},
]{doclicense}

\usepackage{doi}
\usepackage[a4paper, margin=1in]{geometry}
\usepackage{xspace}
\usepackage{typed-checklist}
\usepackage{orcidlink}

\newcommand\rfcword[1]{\emph{#1}\xspace}
\newcommand\must{\rfcword{must}}
\newcommand\mustnot{\rfcword{must not}}
\newcommand\required{\rfcword{required}}
\newcommand\shall{\rfcword{shall}}
\newcommand\shallnot{\rfcword{shall not}}
\newcommand\should{\rfcword{should}}
\newcommand\shouldnot{\rfcword{should not}}
\newcommand\recommended{\rfcword{recommended}}
\newcommand\may{\rfcword{may}}
\newcommand\optional{\rfcword{optional}}

\newcommand\filename[1]{\texttt{#1}\xspace}
\newcommand\readme{\filename{README}}
\newcommand\authorxml{\filename{author.xml}}
\newcommand\program[1]{\textsc{#1}\xspace}

\newcommand{\orcidauthorBENNETT}{0000-0002-1678-6701}

\newcommand\blfootnote[1]{%
  \begingroup
  \renewcommand\thefootnote{}\footnote{#1}%
  \addtocounter{footnote}{-1}%
  \endgroup
}

\title{The TELOS Collaboration Approach to Reproducibility and Open Science}

\author{
  Ed Bennett\,\orcidlink{\orcidauthorBENNETT} \\
  \href{mailto:e.j.bennett@swansea.ac.uk}{\texttt{e.j.bennett@swansea.ac.uk}} \\
  {\small Swansea Academy of Advanced Computing, Swansea University, Fabian Way, Swansea, United Kingdom}
  \\\\
  \href{https://telos-collaboration.github.io}{for the TELOS Collaboration} \vspace{4pt} \\
  \href{https://telos-collaboration.github.io}{\includegraphics[width=2cm]{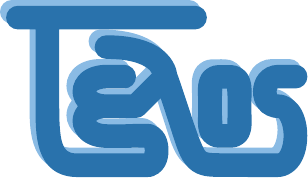}}
}

\date{2025-04-02, version 0.1.1}

\begin{document}

\maketitle

\abstract{
  The TELOS Collaboration is committed to
  producing and analysing lattice data reproducibly,
  and sharing its research openly.
  In this document,
  we set out the ways that we make this happen,
  where there is scope for improvement,
  and how we plan to achieve this.
  This is intended to work both as a statement of policy,
  and a guide to practice for those beginning to work with us.
  Some details and recommendations are specific to
  the context in which the Collaboration works
  (such as references to requirements imposed by funders in the United Kingdom);
  however,
  most recommendations may serve as a template for
  other collaborations looking to make their own work reproducible.
  Full tutorials on every aspect of reproducibility are
  beyond the scope of this document,
  but we refer to other resources for further information.
\blfootnote{\doclicenseThis}
}

\makeatletter
\renewcommand\tableofcontents{%
    \subsection{\contentsname
        \@mkboth{%
           \MakeUppercase\contentsname}{\MakeUppercase\contentsname}}%
    \@starttoc{toc}%
  }
\makeatother

\section{About this document}

\subsection{Introduction}

Reproducibility and openness are becoming increasingly important in science.
Many fields have experienced some form of replication or reproducibility crisis,
where work that had been taken as fact was subsequently found
to be based on shaky results that did not stand up to scrutiny.
As a fully computational discipline,
there is no reason in principle
that work in lattice quantum field theory should not be fully reproducible end to end.
In some cases,
this comes into tension with making optimal use of computing resources;
in this document we will make explicit what compromises we make,
and where improvements could be made to reduce the impact of these.
It is also important that limited public resources not be wasted
repeating the same work in multiple contexts
because different groups did not have access to each others' data.
Our aim is to push forward the boundaries of human knowledge,
and the main barriers to this are human and computational capacity;
we believe that maximising the amount of our work that we share,
and the ability of others to access and build on top of it,
is the optimal way to overcome these.

The TELOS Collaboration~\cite{telos} undertakes
Theoretical Explorations on the Lattice with Orthogonal and Symplectic Groups.
This document is in part a statement of the way we currently work,
in part an indication of the way we intend to work moving forward,
and in part a guide to how to do this.
Where it makes recommendations or imposes requirements,
these are to and on work being performed by the TELOS Collaboration;
however,
it is our hope that others will draw inspiration from these guidelines
and choose to adopt similar ones for their own work.
For space reasons,
it cannot be a complete guide to every aspect of this process;
it instead links to additional material to learn more.

If you find the recommendations presented here useful in your work,
we would ask that you cite them in publications that have relied on them.
If you wish to adapt this document for your own context,
you may do so compatibly with the Creative Commons Attribution License,
for example,
by including a citation back to this document.

We welcome constructive suggestions for how we can improve our practice,
for tools that may help us to achieve our objectives more easily,
and for how this document could be made clearer and more helpful,
although we do not commit to adopt every good suggestion we receive.

\tableofcontents

\subsection{How to read this document}

The remainder of this document is structured as follows:
in Section~\ref{sec:definitions},
we define a number of relevant terms we will rely on throughout the discussion.
In Section~\ref{sec:publications} we discuss open-access publications,
and some other details of the publication process.
In Section~\ref{sec:data} we discuss how and why we publish our data.
In Section~\ref{sec:workflows} we discuss our approach to developing and sharing
the workflows used to analyse those data.
In Section~\ref{sec:hpc-software} we outline our aspirations to
better enable reproducibility of our HPC-based computations.
Appendix~\ref{app:checklist} is provided for reference,
and also as a summary of the main requirements discussed in the main body.

This document contains relatively densely-packed information,
and many of the concepts discussed are tightly linked to each other.
It is ordered to give a relatively logical progression;
however,
the reasoning behind the recommendations in some sections
may make more sense in the context of having read later ones.
As such,
the interested lattice-oriented reader is invited to first skim-read the document
from Section~\ref{sec:definitions} to Section~\ref{sec:hpc-software},
before going back and re-reading these sections in more detail.

While the primary focus of the guidance is
numerical research using lattice quantum field theory techniques;
we briefly discuss the applicability of the guidance to other related fields
in Appendix~\ref{app:nonlattice}.
An interested non-lattice reader might then read Section~\ref{sec:definitions},
then Appendix~\ref{app:nonlattice},
and then the sections outlined in Appendix~\ref{app:nonlattice}.

\subsection{Version history}

These guidelines are intended as a living document,
evolving as our practice develops
and as we progress on achieving
some of the aims that are currently aspirational or works in progress.
The canonical working version of the document
is available from GitHub~\cite{this-github};
more significant updates
will also be posted to the arXiv and to Zenodo~\cite{this-zenodo}.

\begin{description}
  \item[v0.1.1 2025-04-02]
        Explain briefly what the TELOS collaboration is and does.
        Clarify what to do if a journal says no to the rights retention statement.
        Clarify some phrasing.
        Avoid using words reserved in Section~\ref{sec:keywords}
        outside of those definitions.
        Make Data and Software availability statements comply with this guidance.
        Recommend \verb|---Analysis workflow| rather than \verb|---Workflow|
        for workflow names,
        consistent with existing work.
        Software availability statement in checklist.
  \item[v0.1.0 2025-04-02]
        First version shared publicly via arXiv and Zenodo.
        Describe version history.
        Tidy up front matter.
        Make license explicit.
        Add Acknowledgments and other end matter.
        Expand Introduction.
        Rename \program{HEmCee} to \program{hmcdj}.
        Fix some typos.
  \item[v0.0.6 2025-03-25]
        Make Appendix B less roundabout.
  \item[v0.0.5 2025-02-10]
        Add dates to version history.
        Make more explicit statement around who the document is for.
        Add reference to template repository for workflows.
        Add appendix discussing applications outside lattice.
        Mention how to validate against the schema,
        and require this.
        Mention software and data availability statements.
        Correct subsubsections on testing and publishing to being subsections.
        Add reading guide.
        Fix some typos.
  \item[v0.0.4 2025-01-28]
        Incorporate changes from first internal review.
        Clarify our lack of legal qualifications,
        and some other language.
        Explicitly state that UKRI requires CC BY\@.
        Recommend a directory structure.
        Recommend maintaining HDF5 structure consistency.
        Mention Zenodo command-line uploads.
        Avoid overfull hboxes.
        Give an explicit example of code to go in a library.
        Mention existing provenance tooling.
        Mention that repositories can continue being used after a Zenodo release.
        Reformat bibliography to include more links.
  \item[v0.0.3 2024-11-29]
        Add making repo public to checklist.
  \item[v0.0.2 2024-11-01]
        Initial quasi-complete draft.
  \item[v0.0.1 2024-10-31]
        Initial internal draft.
\end{description}

\subsection{Definitions}\label{sec:definitions}

\subsubsection{Reproducibility}\label{sec:reproducibility}

In our work,
we use the definitions of ``reproducibility'' and associated terms
used by the Turing Way project~\cite{the_turing_way_community_2023_7625728}.
Specifically,
we define an analysis as:
\begin{description}
  \item[reproducible]
        when another researcher is able to take the same data,
        perform the same analysis on it,
        and obtain the same results;
  \item[replicable]
        when another researcher is able to produce their own equivalent dataset,
        perform the same analysis on it,
        and obtain the same results;
        and
  \item[robust]
        when another researcher is able to take the same data,
        perform an alternative analysis on it,
        and obtain results leading to the same conclusions.
\end{description}
Reproducibility is the simplest of these to achieve,
and may be considered as trivial;
indeed,
science is predicated on it,
and so work that is not reproducible is of limited value.

\subsubsection{FAIR}

Data and software tools may be described as FAIR if they are
Findable, Accessible, Interoperable, and Reusable~\cite{wilkinson2016fair}.
Specifically,
data and metadata are
\begin{description}
  \item[Findable]
        if they have a unique, persistent identifier,
        are described by rich metadata linked back via this identifier,
        and are indexed or registered in a searchable resource;
  \item[Accessible]
        if they are retrievable using a standardised, open communications protocol,
        allowing for authentication and authorisation where necessary,
        and where metadata remain available even after associated data are not;
  \item[Interoperable]
        if they use a formal, accessible, shared, and broadly applicable language for knowledge representation;
        if they use vocabularies that follow the same FAIR principles,
        and use qualified references to other data and metadata;
        and
  \item[Reusable]
        if they are richly described with accurate and relevant attributes,
        released with a clear and accessible license,
        associated with detailed provenance,
        and meet community-relevant standards.
\end{description}

\subsubsection{Open Science}

Open Science or Open Research is
the movement to make all research accessible to all levels of society.
This may include not just papers,
but also other outputs like data, software, and experimental samples,
as well as including those from outside academia in
the process of conducting research itself.
This can only ever be an ideal to be strived towards,
being tensioned against the costs of opening access to limited resources.

\subsubsection{Copyright and licensing}

\emph{
  In this section we discuss certain aspects of copyright law.
  It is worth mentioning explicitly that
  the authors of this document are not legal professionals,
  and the descriptions below represent our best understanding of the law
  as it applies in the jurisdictions in which we work;
  it is not legal advice.
}

Under copyright law,
the default state is that
when you provide someone with a piece of copyrighted work
or a copy thereof,
they have no right to make further copies of it.
In a digital world,
this includes many operations that are essential for consuming content,
such as downloading a copy of a paper to read.
Some actions may be construed as ``fair dealing'' or ``fair use'',
where there is no penalty,
but many important actions
like reusing or building on top of a piece of work
require explicit permission.
All creative work,
including academic papers, data, and software code,
automatically receives copyright protection at the moment of its creation,
which typically lasts for seventy years beyond the death of its creator.

To avoid each person having to write their own forms of permission
(many of which would be found unenforceable on contact with a lawyer),
certain organisations have defined standard ``copyright licenses''
that may be applied to a piece of work
to grant others specific permissions regarding what they can do with it
and what constraints they have to comply with to benefit from these permissions.

\paragraph{Creative Commons}

Creative Commons defines a number of licenses suitable for papers and data.
These can be recognised as a series of letters starting with ``CC''.
The most common,
and most relevant to our usage,
is the Creative Commons Attribution License,
CC BY~\cite{cc-by}.
This gives recipients the right to make and distribute copies of a work,
provided that the original author is clearly attributed.
This is the license that is mandated by UKRI
(unless there are strong reasons to choose a different license),
who fund our research in the United Kingdom.

\paragraph{Software Licenses}

Creative Commons licenses are not considered
suitable for licensing software source code,
due to concerns around software-specific issues such as patents.
Instead,
it is better to use specific software licenses.
Two common ones are the MIT License~\cite{mit},
which grants rights similar to those granted by CC BY,
and the GNU General Public License~\cite{gpl},
which imposes an additional requirement that
if derivative work is distributed,
this distribution is on the same terms.

Much modern gadgetry makes use of MIT- and similarly-licensed software,
with a long list of attributions,
but likely little support given back to the original authors.
GPL-licensed software is less used in this context,
due to organisations not wanting to share their proprietary software
and potentially give assistance to their competitors.
The aim of GPL is that
anyone is free to modify and improve the software on their own devices;
however,
the extent to which this has succeeded is limited.

\subsubsection{Paper stages}

A paper written by this collaboration
typically has a number of stages in its preparation process:

\begin{description}
  \item[Draft]
        Version worked on by one person,
        or shared within the collaboration for comment
  \item[Preprint]
        Version posted on the arXiv for comments from the community,
        in advance of submission to a journal.
  \item[Author Accepted Manuscript]
        Last version submitted to a journal before acceptance.
        Incorporates feedback from peer reviewers.
  \item[Version of Record]
        Version distributed by the journal.
        In some cases
        (for example,
        in the Journal of High Energy Physics
        and Proceedings of Science)
        this will appear near-identical to the Author Accepted Manuscript;
        in others
        (such as in Physical Review D)
        this will have been
        translated by the editorial office into a different style,
        likely with one or more rounds of feedback
        and checking of proofs.
\end{description}

\subsubsection{Persistent identifiers}

A persistent identifier is
an identifier that can be used to refer to
a publication,
a data asset,
or some other object,
that is unique within some system,
and which provides some guarantee that it will remain available in the future.

The Digital Object Identifier,
or DOI,
is an example of a persistent identifier.
DOIs are issued by journals to refer to articles
and by arXiv for preprints,
and can also be generated for data and code assets.

\subsubsection{Keywords}
\label{sec:keywords}

The key words
``\must'',
``\mustnot'',
``\required'',
``\shall'',
``\shallnot'',
``\should'',
``\shouldnot'',
``\recommended'',
``\may'',
and ``\optional''
in this document are to be interpreted as described in RFC 2119~\cite{rfc2119}.

\section{Publications}
\label{sec:publications}

\subsection{Open Access}

There has for a long time been a culture in theoretical particle physics,
including in lattice,
of sharing preprints openly using the arXiv~\cite{ginsparg2021lessons}.
More recently,
the SCOAP3 agreement has meant that
articles published in
typical journals for particle physics,
such as Physics Letters B and Physical Review D,
are made available as open access without additional charge.
However,
since not all journals or topics are included in this agreement,
and to ensure
that readers not able to afford journal subscriptions
are able to access the corrected versions of papers,
our funders now require us to retain rights to
``Author Accepted Manuscripts'',
that is,
the versions of papers as accepted by the journal,
after peer review,
so that they can be shared under an open license.

To enable this,
all papers submitted by the TELOS collaboration
\must include the following text verbatim in the submitted manuscript:

\begin{quote}
  For the purpose of open access,
  the authors have applied a Creative Commons attribution (CC BY) licence
  to any Author Accepted Manuscript version arising.
\end{quote}

This is referred to as ``rights retention'',
and is a requirement from our funders UKRI\@.
This \should be placed in a paragraph titled
``Open Access Statement''
immediately after the Acknowledgments section.
Publishers sometimes remove this statement on publication;
this does not affect the process.
If a publisher requires retraction of this statement for the paper to be processed,
or otherwise attempts to block sharing of the Author Accepted Manuscript,
the publication process \mustnot continue
before the issue is raised and discussed with UKRI\@.

We \must also upload either a CC BY licensed Author Accepted Manuscript,
or a CC BY licensed Version of Record,
to an institutional repository.
Where an article has not been covered by SCOAP3,
we \mustnot upload a non-CC BY licensed Version of Record;
the Author Accepted Manuscript \must be uploaded in that case.

\subsection{\authorxml}

When the INSPIRE database service ingests papers from arXiv,
it typically has to take the plain-text author list
and parse it into a structured collection of references to known or new authors.
This process is error-prone,
particularly for larger collaborations,
and for individuals whose names collide with common words associated with authorship.
(For example,
``Ed'' is both a name
and an abbreviation that may be placed in front of a name
to indicate that the latter's owner is an ``editor''.
This means that people with the former name frequently have their names
incorrectly recorded on INSPIRE.)

To avoid needing to manually send corrections to INSPIRE after each publication,
INSPIRE recommends providing
a structured, machine-readable set of metadata regarding authorship of a paper.
This \should be done in the form of a file named \authorxml,
having a structure described in Ref.~\cite{inspire-authorxml}.

A template for this file is available in the TELOS Collaboration Resources repository~\cite{resources}.
If this is used,
elements marked \verb|TODO| \must be replaced,
and authors' details \must be checked,
before publication.

When preparing to submit to the arXiv,
in addition to the LaTeX sources and assets,
two additional files \should be included:

\begin{itemize}
  \item
        The file \authorxml,
        prepared as described above.
  \item
        The file \filename{author.dtd},
        a schema defining the structure that \authorxml follows,
        available from Ref.~\cite{inspire-authorxml}.
\end{itemize}

The file \authorxml \must be validated for correctness against the schema,
by following the instructions in the documentation at Ref.~\cite{inspire-authorxml}.

\subsection{Acknowledgments}\label{sec:acknowledge}

It bears reminding that our work is enabled by
the work and contributions of many others outside our collaboration,
who \must be appropriately acknowledged in our work.

This includes:

\begin{itemize}
  \item
        We \must specify what software was used to perform the work,
        both so that the original authors can gain credit,
        and so that others know which software to use to reproduce our work.
        This \must be done in the Software Availability statement,
        discussed in Sec.~\ref{sec:statements}.
        Additionally,
        this \may be done
        in the section of the narrative where the algorithms and tools are introduced.
  \item
        We \must specify what computing resources were used to perform the work,
        to comply with our obligations to the HPC facilities,
        so that we remain able to access them in the future.
  \item
        Where we have used open data,
        we \must cite the original datasets.
  \item
        We \must acknowledge data storage resources where they have been used.
  \item
        We \must follow all other standard non-computation-specific acknowledgment practices,
        including acknowledging our funders,
        and those who have contributing to the work whilst not being authors.
\end{itemize}

\subsection{Additional statements}
\label{sec:statements}

It might be difficult to fluidly incorporate
full information around some of the points discussed above into the main text,
or such information might accidentally be removed during editing,
or a reader might not easily find it when they look for it.
To be maximally explicit,
a paper \must include the following two paragraphs,
after the Open Access Statement:

\begin{itemize}
  \item
        A paragraph titled ``Software Availability Statement'',
        listing and citing all relevant software used,
        including version information,
        and a brief description of what each was used for.
        This \must include both software used for data generation on HPC,
        and software used for subsequent data analysis
  \item
        A paragraph titled ``Data Availability Statement'',
        listing and citing all datasets used.
        This \must include both open data used
        (both our previous work,
        and the work of others),
        and new data generated for this work
        (which \must be published,
        as discussed in Section~\ref{sec:data}).
\end{itemize}

\section{Data}\label{sec:data}

The work that we present in publications can represent
the output of hundreds of thousands of pounds' worth of computer time,
and terawatt hours of energy.
To ensure that this is not wasted,
and so that others can reproduce our work
and extract additional results beyond our initial work
without needing to spend similar resources to regenerate the data,
it is vital that we share our data openly.
This also enable those quoting our data to do so easily
without needing to transcribe numbers,
which is liable to introduce errors.

Every peer-reviewed article that presents new data
\must have an associated data release,
including the new data that were generated in its preparation.
A work that generates no new data,
only presenting previously-analysed data,
\shouldnot have a data release.
Non-peer-reviewed work,
such as conference proceedings,
\may instead refer to
the data release of an upcoming peer-reviewed article
rather than having a separate data release;
however,
if such a work will not be forthcoming,
a data release \should be prepared and published.

A publication \must cite the associated data release using its DOI\@.
Where a publication makes use of data prepared for a previous publication,
the associated data release \must also be cited if it exists.
If the publication did not have an associated data release,
the data used from it \must instead by included in
the current publication's associated workflow release.

The data release associated with a publication
\should be published in advance of the preprint being made available,
and \must be published before the paper is published in the journal.

\subsection{What to include in a data release}

To maximise the utility of a data release to others,
a number of classes of data are needed.

\subsubsection{Final numbers}\label{sec:dr-numbers}

To maximise the utility of our results to those looking to make direct use of them,
data releases \must include all numbers that are plotted as points on a graph,
and \should include all parameters for curves fitted using lattice data
where these are of use to others.
These \must be provided in CSV format,
to maximise ease of use to those who might only be familiar with spreadsheet software.
The number of different CSV files \should be minimised:
data that are characterised by the same parameters
\should be combined into a single file.
For example,
there \may be one file for numbers relating to individual ensembles,
a second file for fit parameters relating to specific values of the coupling,
and a third for fit parameters for continuum limit extrapolations.

Currently there is no community-defined metadata schema for structuring such data.
Nevertheless,
we \should aim to use a common form for our data
to maximise interoperability with others' work.

\begin{itemize}
  \item
        Numbers with single symmetric uncertainties \must be formatted with columns labelled
        \verb|{name}_value| and \verb|{name}_uncertainty|.
  \item
        Numbers with asymmetric uncertainties \must use the suffixes
        \verb|upper_uncertainty| and \verb|lower_uncertainty|.
  \item
        Numbers with multiple sources of uncertainty \must use
        \verb|uncertainty| or \verb|statistical_uncertainty|
        as the suffix for the statistical uncertainty,
        and \must use suffixes of the form
        \verb|{uncertainty_type}_uncertainty|
        for other sources of uncertainty,
        such as systematic.
  \item
        Metrics associated with each other \must use a common prefix.
        For example,
        the mass and matrix element from fitting a pseudoscalar correlation function
        and the associated $\chi^{2}/\textnormal{d.o.f.}$
        might use column names
        \verb|ps_mass_value|,
        \verb|ps_mass_uncertainty|,
        \verb|ps_matrix_element_value|,
        \verb|ps_matrix_element_uncertainty|,
        and \verb|ps_chisquare|.
  \item
        Numbers \must be given to the full machine precision at which they were originally presented,
        not truncated or rounded.
  \item
        Column names \should be documented in the data release documentation
        (see Sec.~\ref{sec:dr-documentation} below).
\end{itemize}

Note that these constraints apply to data files within the data release,
not to numbers presented in the paper
(for example,
in tables).
For example,
tables in papers \shouldnot present numbers to machine precision in almost all cases.

\subsubsection{Input files}

To enable others to reproduce our work on HPC if they wish to,
a data release \may include the raw input files,
if they have been retained in a form that reproduces the original work.
If this is done,
the release \should also include sufficient documentation
to allow a competent practitioner to be able to use the input files to reproduce the data.

\subsubsection{Data from HPC}\label{sec:dr-from-hpc}

The TELOS Collaboration's workflow is typically that
some number of gauge ensembles are generated using an HMC-like algorithm,
taking significant computational resources
and generating substantial data.
Then,
observables are computed on each configuration in each ensemble,
still requiring HPC,
but more modest resources,
and generating outputs files small enough to process on a workstation.
These are then transferred off HPC for final statistical analysis.

Given the constraints of storage,
and to enable analysis workflows to be reproducible,
our data releases \must
share the observable results that are the inputs to the statistical analysis.
This \should be in as close to its native form as possible;
to date,
for most of our work,
this is text-based log files from \program{HiRep} and \program{Grid}.
Where necessary for filesize reasons,
these \may be thinned by removing redundant, repeated log lines.

To make the download process simpler for a reader only interested in a specific observable,
these \may be structured so that
each observable or class of observable
is in a separate directory hierarchy.
If doing so,
it is convenient to maintain the same structure for both
the files on the HPC facility
and the files used locally for analysis,
so that the files can be transferred off HPC with minimal reorganisation necessary.
The following directory structure \may be used:
\\
\texttt{\emph{<Gauge Group>}/n\emph{<Flavour 1 representation><Flavour 1 count>}[\_n\emph{<Flavour 2 representation><Flavour 2 count>}]/beta\emph{<beta>}/m\emph{<Flavour 1 representation><Flavour 1 mass>}[\_m\emph{<Flavour 2 representation><Flavour 2 mass>}]/\emph{<NT>}x\emph{<NX>}x\emph{<NY>}x\emph{<NZ>}}\\
for example,
for $\mathrm{Sp}(4)$ with fermions in two representations,
this might become\\
\texttt{Sp4/nF2\_nAS3/beta7.2/mF-0.72\_mAS-1.14/64x48x48x48}.

\subsubsection{Repackaged data}

Since it is tedious to write large amounts of code to parse data from log files,
and the reading process itself is relatively slow,
if the original output was in a text-based format,
a data release \should also include the input data to the statistical analysis
as discussed in Sec.~\ref{sec:dr-from-hpc} above,
reformatted into a packed format readable by standard libraries.

\begin{itemize}
  \item
        The HDF5 format~\cite{hdf5} \should be used for these data.
  \item
        A single HDF5 file for all ensembles \should be preferred
        to one file per ensemble per measurement,
        to minimise the effort needed to download and use the data.
  \item
        Datasets and groups in the HDF5 structure \must be given
        sufficient metadata to identify the specific ensembles they refer to,
        and what specific observables were computed and parameters used to do so.
  \item
        We are currently working on identifying
        structures for HDF5 files that enable more efficient compression to be achieved.
        Until this work completes,
        the HDF5 file structure \should be consistent with
        that used for Ref.~\cite{bennett_2024_13819562}.
\end{itemize}

\subsubsection{Metadata and analysis parameters}

As will be discussed in Section~\ref{sec:workflows} below,
to ensure that others are able to reproduce our work,
it is vital to share the workflows that are used to analyse our data.
These workflows will typically rely on knowing information about the data,
available in a more compact and quickly-readable form than
looking at the headers of each data file separately,
and will further rely on information that we choose as part of our analysis.
For example,
which ensembles do we consider for a particular analysis,
or where to we judge the plateau region of an observable to be for each ensemble.

This information is still data,
so belongs in the data release rather than being hidden mixed into the analysis workflow.

\begin{itemize}
  \item
        A data release \should contain one or more files containing metadata,
        and if appropriate,
        analysis parameters,
        for the analysis workflow to use as input.
  \item
        Metadata and analysis parameters \must be provided in a plain-text,
        human-readable format.
        This \may be a tabular format such as CSV,
        or a more structured format like YAML\@.
  \item
        Metadata and analysis parameters \should be consolidated into as few files as is reasonable,
        similarly to output numbers discussed in Section~\ref{sec:dr-numbers}.
  \item
        The structure of each metadata/parameter file,
        for example,
        field or column names,
        \should be documented similarly to those for output numbers
        discussed in Section~\ref{sec:dr-numbers}.
  \item
        Metadata/parameter files \may be prepared using tools;
        for example,
        it is more convenient to prepare a large CSV file using a spreadsheet program
        rather than by hand in a text editor.
\end{itemize}

\subsubsection{Documentation}\label{sec:dr-documentation}

No matter how careful we are to construct our data releases carefully,
we will not be able to remove all ambiguity while maintaining a usefully compact release.
As such,
it is always necessary write documentation to enable others to understand our releases.

The TELOS Collaboration provides data release documentation in a file called \filename{README.md}
included in the data release.

\begin{itemize}
  \item
        A data release \must contain one or more files containing documentation.
  \item
        The primary documentation \should have a filename beginning \readme.
  \item
        Documentation \must be provided in a plain-text, human-readable format.
        At time of writing,
        this \should be Markdown.
  \item
        Where multiple documentation files are provided,
        each \must be cross-referenced from the main \readme.
  \item
        The main \readme \must provide the following elements:
        \begin{itemize}
          \item
                Information on the work the data were produced to enable,
                including a link to the article(s) in which they were presented.
          \item
                A listing of the files included in the release,
                and a description of what each file contains.
        \end{itemize}
  \item
        Where multiple files (or directories) are grouped into archive files,
        these archives \should contain their own \readme files
        documenting the data formats used for the files therein.
        Such files \must be referred to from the main \readme file.
  \item
        Descriptions of data formats, columns, etc.,
        \should be recycled from previous releases where appropriate.
        (If a release breaks compatibility with the format documented for a previous one,
        it \should be carefully considered whether this is necessary and justified,
        and \should be explicitly noted in the documentation.)
  \item
        Where single files are included in the data release,
        their structure \should be documented in the main \readme file.
  \item
        Portions of the main \readme file \may be used as
        the basis of the data release description.
\end{itemize}

\paragraph{Markdown}

For an introduction on how to write Markdown,
see for example Ref.~\cite{markdown-guide}.
For information on including mathematical expressions in Markdown documents,
see for example Ref.~\cite{github-markdown}.

\subsection{Where and how to publish a data release}\label{sec:publish-data}

\subsubsection{Where to publish}\label{sec:where}

Data releases \must be published using a platform that provides a persistent identifier
(preferably a DOI),
and that commits to maintaining availability of the dataset for an appropriate period.

At time of writing,
the TELOS Collaboration publishes its data releases using Zenodo~\cite{zenodo}.
Zenodo provides 50GiB of storage per dataset by default,
and allows up to 100 files per dataset.
(Zenodo can exceptionally grant up to 200GiB per dataset on a case-by-case basis.)
A release \mustnot be split into many Zenodo datasets in order to bypass this limit.

\subsubsection{Obtaining a DOI before data are ready}\label{sec:get-doi}

It is frequently desirable to have a DOI to put into a manuscript
before the data are ready to be uploaded,
either to avoid having a ``TODO'' item in the draft,
or because the data release is not ready at time of sharing a preprint.
In particular,
if a preprint is being released in advance of the data release being ready,
it \should refer to the DOI at which the data will ultimately be made available.

To obtain a DOI from Zenodo before the data are ready:

\begin{itemize}
  \item
        Start a new upload in Zenodo.
  \item
        Under ``Basic Information'', answer ``No'' to the question
        ``Do you already have a DOI for this upload?''.
  \item
        Select ``Get a DOI now!''
  \item
        Note the DOI that is generated,
        and add it to the references of the manuscript.
  \item
        Add a placeholder file
        (for example, the draft dataset \readme)
        and an author to the draft release.
  \item
        Click ``Save draft''.
\end{itemize}

The saved draft can then be used as the basis for the full release once it is ready,
and citations to it from the draft or preprint will then be correctly resolved.

\subsubsection{Structuring the release}

Owing to the limitation on the number of files in a release,
and the lack of directory structure in Zenodo,
some files need to be packaged into archives.
However,
to maximise the utility of the release to
those who might only want a small part of it,
thought is needed as to what to group together.

\begin{description}
  \item[The \readme]
        \must be a separate file.
  \item[CSV files]
        \should be uploaded as individual files,
        such that someone wanting to quote a datum from one of them
        need not download a lot of irrelevant files.
  \item[Raw data]
        \should be packaged into one or more archives.
        If there are multiple classes of data
        (for example,
        HMC logs, gradient flow logs, and correlation function logs),
        then these \may be packaged separately.
        Each archive \should produce a \filename{raw\_data} directory,
        containing a subdirectory for the specific observable,
        matching the structure expected by the analysis workflow.
        As discussed above,
        each archive \should contain its own \readme.
  \item[Repackaged raw data]
        \should be uploaded as a single large HDF5 file.
  \item[Metadata and analysis parameters]
        \may be packaged into one or more archives,
        or uploaded as single files,
        as appropriate.
  \item[Input files]
        \should be packaged into one or more archives.
\end{description}

Where archives are used,
these \should be ZIP files,
not \filename{.tar.gz} files,
since the former can be previewed by Zenodo.
They \should use a high compression level,
to reduce the storage and bandwidth requirements,
in particular for raw data comprising logs,
which are naturally storage-inefficient.

\subsubsection{Completing the upload form}\label{sec:upload-form}

\paragraph{Files}

The files discussed above \must be uploaded.
It is \recommended to upload large files one at a time;
concurrent uploads typically fail,
and the form needs to be reloaded to delete the failed files and re-upload.
For large files,
it is \recommended to use a wired network connection
with a high-bandwidth uplink to the Internet.
For very large files,
the web interface can struggle to complete an upload even with a high-bandwidth uplink.
In this case,
it is \recommended to use command-line tools to upload these files,
such as \verb|zenodo-upload|~\cite{zenodo-upload}.

\paragraph{Basic information}

\begin{itemize}
  \item
        The DOI \should have been pre-registered per the above.
  \item
        The Resource Type \should be ``Dataset''.
  \item
        The Title \should be the title of the corresponding article,
        followed by ``---Data release''.
        Where the release supports multiple concurrent articles,
        the Title \may use both article's titles,
        separated by ``and''.''
  \item
        The Creators \must comprise all authors of the corresponding article(s).
        Unless otherwise agreed,
        all authors \should be listed in the role of ``Project member''.
  \item
        The Description \should contain a summary of the \readme,
        linking to the relevant article.
        For example,
\begin{verbatim}
This release contains all data generated in preparing the paper
"[paper title]" (insert link).
It includes two classes of data:
 - Raw data, as generated from the HMC measurement code running on HPC,
   in their native formats.
 - Metadata around the analysis of the ensembles, in YAML format.
Due to their size, raw gauge configurations are not included in this release.

Further documentation on the directory structure and file formats is provided
in the file README.md
\end{verbatim}
  \item
        An additional Description \must be added,
        comprising the Acknowledgments of the relevant paper.
  \item
        The License \should be specified as the Creative Commons Attribution License
        (CC BY).
\end{itemize}

\paragraph{Recommended information}

This section \may be left blank.

\paragraph{Funding}

Grants from the European Union,
and from United Kingdom research councils,
\must be listed here.
Searching the grant ID surrounded by quotation marks,
for example,
\verb|"EP/V052489/1"|,
typically returns only the grant of interest.
(Omitting the quotation marks results in Zenodo including many irrelevant results.)
Grants from other countries \may be listed here also.

\paragraph{Alternative identifiers}

This section \may be left blank.

\paragraph{Related works}

This \should link to:
\begin{itemize}
  \item
        The arXiv preprint,
        with relation ``is described by''.
  \item
        The published article,
        with relation ``is described by''.
  \item
        The analysis workflow release,
        with relation ``is required by''.
\end{itemize}

\paragraph{References}

This section \may be left blank.

\paragraph{Software}

This section \may be left blank.

\paragraph{Publishing information}

This section \may be left blank.

\paragraph{Conference}

This section \may be left blank.

\paragraph{Domain specific fields}

This section \may be left blank.

\subsection{Field configurations}

Field configurations take significant computational resources to generate,
and contain significantly more information than one person or collaboration can extract.
As such,
we aspire to share our field configurations openly,
to maximise the benefit that can be obtained from them.
(To this,
we apply the constraint that
configurations \shouldnot be retained if the cost of storage outweighs
the cost of regenerating them given the input parameters.)
In principle,
this would also allow data from different collaborations
to be analysed many of the same systematics,
rather than comparing data that have been computed and analysed with different approximations.

The International Lattice Data Grid (ILDG)~\cite{ildg-organization}
defines standards for interoperability and sharing of field configurations.
We aim to share our configurations using the UK Regional Grid of the ILDG\@.
Currently this is not in service;
further information on preparing and pushing configurations to this service
will be provided once it is available.

In the interim,
when generating ensembles requiring non-trivial computational effort,
we \must retain sufficient information that
the ILDG configuration and ensemble metadata \may be completed at a later date.
This \should include:

\begin{itemize}
  \item
        The gauge group
  \item
        The gauge action, associated parameters (e.g. $\beta$), and boundary conditions
  \item
        The fermion actions, associated parameters (e.g.\ bare mass), representations, and boundary conditions
  \item
        The algorithm used to perform the generation
  \item
        The software (including version information) used to perform the generation
  \item
        The precision worked at by the software used to perform the generation
  \item
        The name and affiliation of the person who performed the generation
  \item
        The date and time at which each configuration was generated
\end{itemize}

\section{Workflows}\label{sec:workflows}

In our data release,
we share data from various stages of our computation.
However,
we \must also share how data get from one stage to the other.
The Royal Society shortly after its founding in 1660
chose as its motto ``\emph{nullius in verba}''---take nobody's word for it---
to signify the importance of this.
It marked a transition to science being based on reproducibility,
where no result could be accepted until others were able to show it for themselves.
This laid the foundation for the growth in science over the past centuries.

In principle,
sharing the data analysis code used to go from input to output data
is not a necessary condition for reproducibility.
A sufficiently precise,
and accurate,
narrative publication
could achieve the same result.
In practice,
however,
specifying the level of necessary detail
and keeping the actual software used synchronised with this
is beyond the capabilities of most authors;
in many fields this has led to a ``replication crisis''
where key findings in a discipline were discovered to be unfounded,
as the description of the methodology did not match what was actually done,
due either to human error in the analysis process,
or imprecise language in the description.

As such,
we assert that codifying the analysis performed in data and software,
and publishing both of these openly alongside narrative articles,
is the \emph{easiest} way to achieve reproducibility of our work.
Others looking to apply our techniques
(as we would hope they would,
if we are doing our jobs properly)
can make use of our code directly,
or can reimplement based on our descriptions,
but cross-check against our implementation for any discrepancies.

Working reproducibly has an initial learning curve,
and requires some more up-front setup than diving straight in to
fiddling with data and producing plots by hand.
However,
it pays substantial dividends as the quantity of data you work with increases,
the workflows become more complex and interlinked,
and the number of projects you work on grows.

In this section we will discuss some of the approaches that we adopt.
This will necessarily be incomplete,
and our approach is likely to evolve over time,
as experience shows us better ways of working,
and new tools and technologies become available.

\subsection{Workflow essentials}
\label{sec:wf-essentials}

All workflows \must be developed under version control.
In the TELOS Collaboration we use Git for this purpose;
if you are unfamiliar with Git,
good introductory guides include Refs.~\cite{swc-git,chacon2014pro}.
Workflows under development \should be regularly synchronised with a hosting service,
to avoid the only copy being on a laptop that might suffer data loss.
We make use of GitHub for this purpose.
Workflows \may be developed under the \texttt{telos-collaboration} organisation;
if not,
they \must be transferred to this organisation before publishing.

Each workflow repository \must contain the following files at root level:

\begin{itemize}
  \item A \readme file, for example \filename{README.md},
  \item A \filename{LICENSE} file, and
  \item A \filename{CITATION.cff} file.
\end{itemize}

It \should also contain the following:

\begin{itemize}
  \item A \filename{.gitignore} file, and
  \item A \filename{.pre-commit-config.yaml} file.
\end{itemize}

\subsubsection{\readme}

The \readme of a workflow release \must contain:

\begin{itemize}
  \item
        A brief description of the workflow,
        including a link to the article presenting the workflow's output.
        This \should be in the form of a DOI\@.
  \item
        A list of prerequisite software needed to be able to use the workflow.
        This is likely to include
        Conda,
        Snakemake
        (which \may be installed via Conda),
        and a LaTeX distribution.
        This \should note any ``gotchas'' that might trip up a potential user,
        but \shouldnot give detailed installation instructions
        that would duplicate the documentation of the software in question.
  \item
        Instructions on setting up the workflow.
        This \should include how to clone the repository,
        what data to download from the data release,
        and where to put it.
  \item
        Instructions on running the workflow.
        This \should be a single command,
        and \must be equivalent to the command that was run
        to generate the assets presented in the corresponding article.
        (You \may,
        for example,
        change the parallelisation options to Snakemake is allowed,
        but \mustnot use an entirely different script.)
  \item
        Information on the approximate expected runtime of the workflow,
        and the machine specification on which this estimate is based.
        (It is useful for a reader to know in advance
        whether they will need to allocate
        minutes,
        hours,
        or days
        to the computation.)
  \item
        Information on where the output of the workflow is placed.
\end{itemize}

Additionally,
the \readme \should contain:

\begin{itemize}
  \item
        The DOI of the workflow;
        this \may be in the form of a DOI badge.
        To obtain a DOI from Zenodo prior to releasing it,
        see the instructions in Section~\ref{sec:get-doi}.
  \item
        A discussion of the reusability of the workflow.
        How much of the workflow is expected to be applicable to other contexts,
        and what work would be needed to do so.
\end{itemize}

\subsubsection{\filename{LICENSE}}

The \filename{LICENSE} file \should contain
the full text of the license under which the workflow is made available.
The TELOS Collaboration uses the GNU General Public License~\cite{gpl} by default.
If there are reasons to prefer another license,
this \must be discussed with the collaboration before applying it.

\subsubsection{\filename{CITATION.cff}}

To aid tools in generating appropriate citations for our work,
we provide metadata on the release in the form of a \filename{CITATION.cff} file.
This is written in the Citation File Format (CFF)~\cite{cff}.
CFF files \may be generated using the \program{cffinit}~\cite{cff-init} tool,
or \may be based on the skeleton example in the resources repository~\cite{resources}.
If the latter is used,
elements marked \verb|TODO| \must be replaced,
and authors' details \must be checked,
before publication.

\subsubsection{\filename{.gitignore}}

The \filename{.gitignore} file is a standard mechanism to tell Git
what files are
(typically)
not wanted to be committed to a repository.
This \may be based on a template such as that provided by GitHub~\cite{gitignore-python},
but \should also specify the directories expected to contain
data downloaded from the data release,
and files output by the workflow.
A minimal \filename{.gitignore} file for a workflow matching the description below might be,

\begin{verbatim}
# Common temporary files
scratch
.DS_Store
.#*
*#*#
*~
*__pycache__
**.pyc
**.ipynb_checkpoints
*.pdf
cache/
tmp/

# Input and output data
.snakemake/
raw_data/
intermediary_data/
data_assets/
assets/
metadata/
!*/.git_keep
\end{verbatim}

Including data files in the \filename{.gitignore} file early in the development process,
even before starting to test the workflow with data,
is important,
because if data files are accidentally committed to the repository,
then everyone who downloads the repository will need to download them:
even if a later commit removes the files again,
they will still be present in the history.
(It is possible to rewrite history to remove unwanted files,
but this takes additional work and requires coordination to not reintroduce them.)

\subsubsection{\filename{.pre-commit-config.yaml}}

We aim for our releases to be useful for others,
and to minimise the work needed when we are developing them.
As such,
we try to make our code easy to read;
it is difficult enough to read one's own code
after a few weeks of not looking at it,
let alone someone else's.
To this end,
we use some basic automated code quality checks,
to keep the basic formatting of code consistent.

For this,
we make use of \program{pre-commit}~\cite{pre-commit}.
When inside a repository,
with \program{pre-commit} available in the \texttt{\$PATH},
running the command
\begin{verbatim}
pre-commit install
\end{verbatim}
adds a hook to the repository that is run when a commit is attempted;
it reads and runs the checks defined in \filename{.pre-commit-config.yaml},
and if any checks do not pass,
the commit is blocked.
Some checks automatically fix the files such that you can retry the commit quickly;
other checks might require manual fixes to be made.

A minimal \filename{.pre-commit-config.yaml} for our work might be:

\begin{verbatim}
default_language_version:
    python: python3.12
repos:
- repo: https://github.com/astral-sh/ruff-pre-commit
  rev: v0.5.1
  hooks:
    - id: ruff
      # Lint rules suggested by ruff docs, just for starters:
      args: [--fix]
    - id: ruff-format
- repo: https://github.com/pre-commit/pre-commit-hooks
  rev: v4.6.0
  hooks:
    - id: check-yaml
    - id: end-of-file-fixer
    - id: trailing-whitespace
    - id: check-toml
    - id: mixed-line-ending
- repo: https://github.com/jumanjihouse/pre-commit-hooks
  rev: 3.0.0
  hooks:
    - id: markdownlint
      files: "content/"
\end{verbatim}

The versions specified are correct at time of writing,
but are likely out of date by the time you read this document.
You \may use \program{pre-commit CI}~\cite{pre-commit-ci} to keep these up to date.

\subsection{Workflow management}\label{sec:wf-manager}

Performing an analysis in lattice quantum field theory has many moving parts.
We typically work with many ensembles,
for each of which we compute many observables.
We want to combine subsets of each of these in different ways,
and perform various fits of them.
Each analysis step takes a different amount of time,
and depends on others having been previously completed.

As discussed above,
we require that a single command be able to reproduce the full analysis,
end to end.
One way we might start thinking about that is by writing a shell script,
to run the various tools needed in order.
We might also consider using a single Python script
that calls the various functions that we need to run,
using subprocesses where we use non-Python tools.
However,
this comes with some limitations:
if we want to run multiple steps in parallel
(as we likely want to,
as the number of steps grows),
we need to manually specify how and where to parallelise.
And if we want to avoid recomputing
(potentially expensive)
steps that we have already completed
and whose inputs have not changed,
we need to manually perform these checks.

To avoid reinventing the wheel,
we instead choose to build on the work of others who have already solved these problems.
A workflow manager is a tool that allows you to specify
a set of rules for how to compute specific outputs given specific inputs;
when given a specific target to generate,
it identifies what steps needs to be performed,
and performs them.
A typical workflow manager does this
by building a directed acyclic graph
(DAG)
of the steps;
this then allows it to parallelise over all steps that do not directly depend on each other.

After a survey of the available options,
we identified \program{Snakemake}~\cite{molder2021sustainable} as a good fit for our typical needs.
It has a number of features that are useful to us:

\begin{itemize}
  \item
        It can run on one or multiple CPU cores.
  \item
        It can manage the software environments needed for running rules using Conda.
        No manual installation of environments is necessary for a user of the release,
        beyond installing Snakemake.
  \item
        It tracks what outputs it has generated using what inputs,
        so only re-runs rules when the input data change.
  \item
        The syntax for rule definitions is built on top of Python,
        so is extensible if the standard syntax does not meet our needs
\end{itemize}

It can also run parts of a workflow on external HPC resources;
this is not a feature we currently use,
but might be useful in future.

Currently,
there are no Snakemake tutorials specific to the lattice context;
while it is our ambition to change this,
for the time being,
we refer to e.g.\ Refs.~\cite{carpentries-snakemake,snakemake-tutorial}.

\subsubsection{Structuring a workflow}

The benefits of a workflow manager discussed above are maximised when
the structure of the workflow is aligned with the strengths of the workflow manager.
Put shortly:
\begin{itemize}
  \item
        The scope of each tool
        (or rule)
        \should be kept as small as possible,
        with the complexity being encoded as relationships between rules in the workflow definition.
  \item
        Parameters \should be known to the workflow manager,
        rather than passing a file containing them.
  \item
        Input and output files \should be specified as command-line parameters,
        rather than being hard-coded.
  \item
        The workflow \shouldnot append to files,
        only create them afresh.
        If joining files together is needed,
        that \should be its own rule.
\end{itemize}

To take a concrete example,
if we have data files for meson correlation functions and gradient flow histories
at various values of the bare mass $am_{0}$,
and we wish to plot the masses of the pseudoscalar and vector mesons
$aM_{\mathrm{PS}}$ and $aM_{\mathrm{V}}$,
normalised by the gradient flow scale $w_{0}/a$
as a function of $am_{0}$,
then the workflow might be structured as the following:

\begin{itemize}
  \item
        A preamble,
        in which the workflow reads a metadata file with information about the ensembles,
        and in particular the plateau positions for each channel for each ensemble.
  \item
        A rule to take one ensemble's gradient flow history,
        and output bootstrap samples for $w_{0}/a$ to a file,
        labelled by an ensemble identifier and the observable name
        (for example,
        \texttt{Sp4nF2b6.7mF-0.62T48L32/\hspace{0pt}w0\_samples.json}).
        This file \should also carry
        metadata and provenance information about the ensemble.
  \item
        A rule to take one ensemble's meson correlation function data,
        and using the plateau position metadata,
        compute the mass of one mesonic channel $aM_{X}$,
        outputting bootstrap samples to a file,
        labelled by an ensemble identifier and the channel name
        (for example,
        \texttt{Sp4nF2b6.7mF-0.62T48L32/\hspace{0pt}ps\_mass\_samples.json}).

        This file \should also carry
        metadata and provenance information about the ensemble.
  \item
        A rule to take the bootstrap samples for $aM_{X}$ and $w_{0}/a$,
        and output the dimensionless product $\hat{M}_{X}=w_{0}M_{X}$,
        to a file labelled by an ensemble identifier and a description of the product
        (for example,
        \texttt{Sp4nF2b6.7mF-0.62T48L32/\hspace{0pt}normalised\_ps\_mass\_samples.json}).
        This file \should also carry
        metadata and provenance information about the ensemble.
  \item
        A rule to take any number of files
        each containing the value of the normalised mass for a single ensemble,
        and a plot style definition,
        and output a plot of the normalised mass against the bare fermion mass
        (the latter obtained from the metadata).
\end{itemize}

To meet the requirements discussed in Section~\ref{sec:data} above,
there would also want to be a rule to take all of the various sample files
and output a CSV that could be included in the data release.

\subsection{Structuring the repository}\label{sec:repository-structure}

We have discussed above the essential files to include in the root directory of the repository.
Now,
let's go into more detail about how to structure the remainder of the repository.

\begin{itemize}
  \item
        Files relating to the definition of the workflow \should be placed in a \filename{workflow/} directory.
        The workflow itself \must be placed in \filename{workflow/Snakefile}.
        Workflows with many rules \should be split into modules;
        these \should be placed in \filename{workflow/rules/}.
        Conda environment definitions \should be placed in \filename{workflow/envs/}.
        This structure matches the standard recommendations for Snakemake projects in general.
  \item
        Source files \should be placed in a \filename{src/} directory.
        For projects of any complexity,
        this \should be structured as a Python package,
        containing an \filename{\_\_init\_\_.py} file,
        so that relative imports can be made.
  \item
        Definitions of,
        for example,
        plot styles \should be placed in a \filename{styles/} directory.
  \item
        Libraries that are not available via standard package repositories
        \should be placed in a \filename{libs/} directory,
        as discussed below.
  \item
        Input data \should be placed in a \filename{raw\_data/} directory.
        This directory \may contain a \filename{.git\_keep} file
        so that it is created when the repository is cloned,
        but all other files in the directory \must be ignored in \filename{.gitignore},
        to avoid accidentally committing large amounts of data.
  \item
        Data that have been quoted from other publications
        that did not provide a data release
        (so where numbers had to be transcribed by hand)
        \should be placed in a \filename{quoted\_data/} directory.
        They \must include documentation as to their provenance,
        including attribution to the original authors.
  \item
        Files containing metadata and analysis parameters
        \should be placed in a \filename{metadata/} directory.
        This directory \may contain a \filename{.git\_keep} file,
        but all other files in the directory \must be ignored in \filename{.gitignore}.
  \item
        Data produced by the workflow but not intended for distribution
        (for example,
        the bootstrap sample files discussed above)
        \should be placed in an \filename{intermediary\_data/} directory.
        This directory \mustnot contain a \filename{.git\_keep} file,
        and \must be ignored in \filename{.gitignore}.
  \item
        Data produced by the workflow and intended for distribution
        (for example,
        CSV files of final numbers to include in the data release)
        \should be placed in a \filename{data\_assets/} directory.
        This directory \may contain a \filename{.git\_keep} file,
        particularly if it is anticipated that users will place files into the directory by hand,
        but all other files in the directory \must be ignored in \filename{.gitignore},
        to avoid accidentally committing large amounts of data.
  \item
        Outputs generated by the workflow for inclusion in a manuscript
        (plots, tables, and variable definitions)
        \should be placed into an \filename{assets/} directory.
        This directory \may contain a \filename{.git\_keep} file,
        but all other files in the directory \must be ignored in \filename{.gitignore}.
        When preparing a manuscript,
        the directory \should be copied directly into the project;
        when re-running the workflow,
        the previous version \should be deleted from the project
        before the updated version is copied in.
\end{itemize}

A template repository with this structure,
including templates for the \required files listed in Section~\ref{sec:wf-essentials},
is provided at Ref.~\cite{workflow-template}.

\subsubsection{Libraries}

For published libraries,
it is typically sufficient to allow Conda to install them from standard repositories,
such as PyPI for Python packages.
For libraries only made available via GitHub or similar forge services,
these \mustnot be specified via a GitHub
(or similar)
URL,
for the same reason we do not use GitHub to publish our workflows---GitHub
does not provide a guarantee of long-term stability.

Instead,
where libraries are internal and/or sufficiently actively developed that
they are not available via a package repository,
these \must be included as Git submodules,
and installed directly from the local copy.
We place local copies of libraries in a \filename{libs} directory.

To download a repository as a submodule,
use the syntax
\begin{verbatim}
mkdir -p libs
cd libs
git submodule add https://github.com/username/reponame
\end{verbatim}
This addition can then be committed to the repository as usual.
Note that the URL used \must be a publicly-accessible HTTPS address,
not a private repository or a \texttt{git@github.com:} SSH address.

To clone a repository that includes submodules,
use
\begin{verbatim}
git clone --recurse-submodules git@github.com:telos-collaboration/workflow_name
\end{verbatim}
When adding a submodule to a repository for the first time,
the \readme \should be updated to include this instruction in place of
the plain \texttt{git clone} that would otherwise be present.

To specify a local copy of a library in a Conda environment specification,
replace the \verb|library_name==0.0.1| or similar specification
generated by \texttt{conda env export}
with \verb|../../libs/library_name|.

Note that when GitHub exports a ZIP file of a repository,
it does not include the contents of any submodules.
As such,
we \must create our own ZIP files of such repositories when uploading to Zenodo.

\subsection{Assets to generate}\label{sec:assets}

In general,
the workflow \should generate four broad classes of output:
plots,
tables,
definitions,
and data assets.

When naming files,
the workflow \mustnot generate files with the same name in different directories.
This is because arXiv only partially supports subdirectories;
while it can compile projects with files in subdirectories,
if two files in different subdirectories have the same name,
neither is accessible to LaTeX,
and the compilation fails.

\subsubsection{Plots}

Plot generation is one of the first tasks researchers approach automating.
A full deep dive into automated plotting is outside the scope of this document,
but we will highlight some key points specific to the way that we work.

\begin{itemize}
  \item
        Plots \should be produced using Matplotlib,
        or a tool based on its engine.
  \item
        Plots \should read in style information from a common style file,
        and \shouldnot have excessive cosmetic overrides.
  \item
        The workflow \should make it simple to change which plot style file is in use,
        for example,
        to switch to using sans-serif fonts and a dark background
        for presentation slides.
  \item
        Plots \should use a standard colour cycle,
        to maximise accessibility for those with colourblindness.
  \item
        Similarly,
        where possible,
        plots \should use differing markers in addition to colours,
        to aid colourblind readers and those with black and white printers
        in interpreting them correctly.
        Where necessary,
        plots \may use markers and colours to differentiate different degrees of freedom.
  \item
        Plots \should use the \verb|layout="constrained"| option to \verb|plt.figure|
        (or to \verb|plt.subplots|)
        to maximise the use of space.
  \item
        Plots \should specify a \verb|figsize|
        equal to the anticipated size in the final manuscript,
        so that they can be used in LaTeX without specifying the \verb|\width| option.
        This ensures that font sizes remain consistent and legible.
  \item
        Plots \should be included in LaTeX files using syntax like:
\begin{verbatim}
\includegraphics{assets/plots/spectrum}
\end{verbatim}
        The \texttt{width=} argument \shouldnot be specified,
        and \shouldnot need to be.
  \item
        Axis and other labels on plots \must match the notation used in the main text.
  \item
        Where the caption of a figure depends on the specific parameters used to generate it,
        then a \filename{.tex} file \should also be generated containing the caption,
        or a definition file generated as described in Section~\ref{sec:definitions}.
        For example,
        if the workflow generates a plot \filename{spectrum\_beta2.3.pdf},
        it would also generate a file \filename{spectrum\_beta2.3\_figure.tex},
        containing something along the lines of
\begin{verbatim}
\begin{figure}
  \includegraphics{assets/plots/spectrum_beta2.3}
  \caption{\label{fig:spectrum-betatwopointthree}
  The spectrum of the theory at $\beta=2.3$}
\end{figure}
\end{verbatim}
        In this instance,
        the figure \should be included in the manuscript using
\begin{verbatim}
\input{assets/plots/spectrum_beta2.3_figure.tex}
\end{verbatim}
  \item
        Plots \should be placed in an \filename{assets/plots/} directory.
\end{itemize}

\subsubsection{Tables}

Tables are frequently constructed by hand,
or generated using tools but then copied and pasted into a manuscript.
This manual intervention is error prone,
and increases the likelihood that the work presented will not be reproducible,
due to numbers from different iterations of the workflow
(or other numbers entirely,
from typographical errors)
being mixed into the final manuscript.

\begin{itemize}
  \item
        There \must be a 1:1 correspondence between tables and \filename{.tex} files generated.
        That is,
        each table \must be in its own \filename{.tex} file.
  \item
        Each table file \must contain the entire table,
        starting with \verb|\begin{tabular}|
        (or equivalent)
        and ending with \verb|\end{tabular}|
        (or equivalent).
  \item
        Where the caption of the table depends on data presented therein,
        the workflow \should also include the caption in the \filename{.tex} file.
        The file \should then contain the entire \verb|table| environment,
        starting with \verb|\begin{table}|
        and ending with \verb|\end{table}|.
  \item
        Table files \must be included in the manuscript using commands along the lines of
\begin{verbatim}
\input{assets/tables/spectrum_table.tex}
\end{verbatim}
  \item
        Tables generated by a single workflow
        \must follow the same formatting conventions,
        in particular around where to place vertical and horizontal lines.
        Where possible,
        this \should be achieved by making use of common functions,
        allowing style changes to be made by changing a single definition.
  \item
        Table definitions \should be preceded by metadata describing them and their provenance.
        This \must be formatted as LaTeX comments
        (i.e.\ lines beginning with \verb|%|).
  \item
        Tables \should be placed in an \filename{assets/tables/} directory.
\end{itemize}

\paragraph{How to generate tables programmatically}

For plain tables of numbers without uncertainties,
Pandas DataFrames have a \verb|styler.to_latex| method to enable this.
For tables of numbers with uncertainties,
this method \may be used in conjunction with
the \program{format\_multiple\_errors}~\cite{fme} library.
Where horizontal lines are needed to break up sections of a table,
currently this is challenging to achieve with Pandas,
but a pull request is open that is hoped to enable this in future.

\subsubsection{Definitions}

Frequently,
we want to discuss numerical results in our papers,
including quoting the numbers themselves.
Our first temptation might be to write the number in the text manually.
However,
if during drafting,
an additional data point is added that changes a fit result in the second decimal place,
it is easy for these numbers to become inconsistent with the actual analysis results.

To avoid this,
our workflows \should output definitions of LaTeX commands
that can be used in documents as placeholders for the numbers,
that will then be filled in by the workflow.
For example,
our workflow might define a command
\begin{verbatim}
\newcommand\PSMassContinuum{0.1234(56)(78)}
\end{verbatim}
which would then be used in the text as
\begin{verbatim}
We find the mass of the pseudoscalar in the continuum limit to be \PSMassContinuum.
\end{verbatim}
which would in turn render in the document as\\
``We find the mass of the pseudoscalar in the continuum limit to be 0.1234(56)(78).''

\begin{itemize}
  \item
        Definitions \should be produced for all numbers presented with uncertainties in a manuscript.
  \item
        Where definitions have been produced,
        they \must be used in place of all instances of the relevant number.
  \item
        Definitions \shouldnot force math mode,
        since in some instances
        they might be used inside a larger mathematical environment.
        (They \may use \verb|\ensuremath|.)
  \item
        Definition files \must have the \filename{.tex} extension.
  \item
        Definition files \should be placed in an \filename{assets/definitions/} directory.
  \item
        Each definition file \should begin with metadata describing the definitions and their provenance.
        This \must be formatted as LaTeX comments
        (i.e.\ lines beginning with \verb|%|).
  \item
        Definition files \must be loaded in the preamble of the manuscript LaTeX document,
        along the lines
\begin{verbatim}
\input{assets/definitions/spectrum_definitions.tex}
\end{verbatim}
\end{itemize}

\subsection{Metadata and provenance tracking}

It is useful
when files are taken outside of their original context
to be able to identify where they came from.
A full realisation of this would require implementing something like PROV~\cite{prov},
as has been done by the authors of Ref.~\cite{Auge:2023vnd};
this is beyond our current capacity,
but \should be on our roadmap.
However,
retaining a subset of provenance information still has value.
This \may be ``stamped'' as a comment in a file,
and/or included as an additional file.

\begin{itemize}
  \item
        When storing intermediary results,
        particularly when JSON or similar formats are used,
        files \should also store structured metadata and provenance information.
  \item
        When producing tables and definition files,
        these \should also include
        unstructured metadata and provenance information
        in LaTeX comments.
  \item
        When producing figures as PDF files,
        including provenance information is non-trivial,
        and is currently considered out of scope.
        When producing other formats
        (for example, Scalable Vector Graphics),
        provenance information \may be included as a comment.
  \item
        When producing directories of output
        (for example, \filename{assets/}),
        there \should be a machine-readable
        (for example, JSON)
        file placed at its root
        indicating metadata and provenance specific to the workflow run that produced it.
\end{itemize}

Things to include in a provenance stamp will vary from case to case,
but \should most likely include:

\begin{itemize}
  \item
        A comment warning that the asset was/assets were automatically generated,
        and \shouldnot be directly modified,
        but instead that the workflow \should be re-run.
        This \should be sorted to appear at the top of the listing.
  \item
        The commit ID of the workflow used to generate the asset,
        and if there were uncommitted changes present.
  \item
        The time at which the asset was generated.
  \item
        The computer using which the asset was generated.
  \item
        The person who ran the workflow.
        (For example, their username.)
  \item
        For assets generated from other data,
        the files that were used as input.
  \item
        Parameters that were used in the generation of the asset.
\end{itemize}

Basic provenance tracking for many of these,
and annotation of output assets with them,
is implemented in Ref.~\cite{bennett_2024_13819431};
this \may be reused rather than reimplementing the functionality from scratch.

\subsection{Standard tools and techniques}

As discussed above,
we make use of Snakemake for workflow management.
Since Snakemake invokes rules using shell commands,
it is relatively simple to coordinate tools written in different languages.
However,
increasing the number of languages increases the complexity of reading and understanding the code,
and of setting up a working environment,
so where possible,
workflows \should use a common language,
and failing that,
as few languages as possible.
In most instances,
new workflow components \should be written in \program{Python}.

Authors of workflows \should reuse existing code where practical,
rather than rewriting a slightly different version.
This helps to maintain consistency between equivalent data presented in different publications.

Where there are common elements used in many workflows,
these \should be grouped into a common library,
to be imported by each workflow.
This avoids having duplicate copies of an analysis in different workflows,
each with slight differences.
For example,
rather than each workflow defining its own method of
solving the Generalised Eigenvalue Problem
(or worse,
different parts of the same workflow but written by different collaborators
having different implementations),
there \should be a common library for the aspects that are common across applications.

Workflows \should aim to be ``DRY''
(``don't repeat yourself'';
as opposed to ``WET'',
``write everything twice'').
Significant amounts of copy-pasting of blocks of code,
or entire files,
\should be avoided;
instead,
common functionality \should be abstracted to a single place.

\subsubsection{Code review}\label{sec:code-review}

Many of the above points require relatively fluency with the language being used,
and some meditation on aspects of research software engineering.
There is no substitute for experience and fluency with a language;
this is gained through experience and feedback.

One way to gain this feedback is through code review.
Where a workflow is being developed collaboratively by multiple people,
changes to the workflow \should be developed in branches,
and then submitted to the main branch as pull requests.
Feedback \should then be given on the content of the pull request
by another co-author of the workflow.
Feedback \must be constructive and positive.
The pull requester and the reviewer \should come to an agreement on
the course of action to take before merging the pull request,
and any identified changes \should be completed before the pull request is merged.

Where the workflow is being developed by a single author,
code review \may instead be done as part of the workflow testing process;
see Sec.~\ref{sec:testing} below.

\subsubsection{Statistics and fits}

Unless there is a strong reason not to,
we make use of bootstrap sampling
to obtain bias-free uncertainty estimates in our work.
(In the majority of our work to date,
we have used 200 bootstrap samples for each analysis.
However,
as we begin performing higher-precision computations with more statistics,
it might be beneficial to use higher numbers of bootstrap samples.)
We make use of
\program{lsqfit}~\cite{lsqfit,peter_lepage_2024_12690493}
and \program{corrfitter}~\cite{corrfitter,peter_lepage_2021_5733391}
for performing fits of data and correlation functions in our work.

\subsubsection{Continuous Integration}

To reduce the amount of data that collaboration members need to have on their own machine,
which has grown larger over time,
and to make it easier to set up a standardised software environment,
we aim to set up a platform
that makes use of Data Version Control~\cite{dvc} or similar
to manage versions of input data for analysis,
and that integrates with GitHub Actions,
to automatically run workflows on centralised infrastructure.

Currently this is still in the early planning phases,
but once ready,
the output from this system \should be used as
the authoritative version of assets to include in publications.
Workflows \should still be tested by other collaboration members to ensure portability and reproducibility.

\subsection{Numerical reproducibility}\label{sec:numerical-repro}

While in Sec.~\ref{sec:reproducibility} we define reproducibility in black and white,
there are in fact certain degrees to which data are reproducible.
They might give compatible results within error bars,
or might give identical results down to the last bit,
or something in between.

An analysis workflow \should give bitwise identical output,
except for differences in headers and other provenance data,
when run on the same platform,
with the same parallelisation options.
It \must give compatible results,
with differences much smaller\footnote{
  We deliberately do not explicitly define ``much smaller'' at this time.
} than the statistical error,
when re-run,
including running on other machines or with different parallelisations.

\subsubsection{Randomness}

Analysis workflows frequently make use of randomness.
In particular,
bootstrap sampling always uses randomness,
and for a finite number of bootstrap samples,
the final numbers obtained will have
an in principle negligible but noticeable dependence on the choice of random numbers.
As such,
we \must fix random seeds before performing a computation with randomness.
This \must be done separately for each workflow rule;
since the workflow can execute in any order compatible with the data dependencies,
different runs might otherwise give different results.

Random seeds \must be different for each ensemble,
since otherwise correlations are introduced between them.
In order to avoid hardcoding specific random seeds,
which could raise questions in the minds of readers of our workflows
(``Where did that number come from?''
``Was it chosen to give a specific outcome?''),
we generate seeds deterministically from input metadata.
To introduce sufficient entropy into these,
it is \recommended to use a hash of the ensemble name as the random seed.
This has the benefit that
having the same bootstrap samples for different observables is simplified.

An example Python implementation might look like:

\begin{verbatim}
import hashlib
import numpy as np

def get_rng(ensemble_name):
    ensemble_hash = hashlib.md5(ensemble_name.encode("utf8")).digest()
    seed = abs(int.from_bytes(filename_hash, "big"))
    return np.random.default_rng(seed)
\end{verbatim}

This approach will give results that are reproducible to near-machine precision.

\subsection{Testing a workflow}\label{sec:testing}

Before a workflow is published,
it \must be tested by a collaboration member and coauthor other than the one who wrote it.
It is \recommended that the tester and workflow author use different platforms,
to identify potential cross-platform issues.
For example,
if the author used macOS to develop the workflow,
it is \recommended for the tester to use GNU/Linux.

The tester \must be given access to only the files that will be submitted to Zenodo and arXiv
(the raw data, metadata and parameters, and output CSVs from the data release,
the source code archive from the analysis workflow release,
and the generated assets that will be included in the article).
In the first instance,
they \mustnot be supervised or guided during the author during this process.
They \must run the workflow end-to-end,
by following the instructions in the workflow \readme,
and verify that the assets
(plots, tables, definitions, and CSVs)
are correctly reproduced.

If difficulties with setting up or running the workflow are encountered,
the tester \may reach out to the author for assistance,
and \should provide feedback to the author.
If the issue is specific to the workflow
(rather than,
for example,
a general difficulty installing Snakemake),
the author \should update the documentation to give more guidance around the difficult aspect.

If the workflow runs,
but does not correctly reproduce the output,
this \must be fed back to the author.
The author \may work with the tester to identify the cause of the discrepancy,
and the divergence \must be fixed before the workflow and data are published.

Once the tester is able to run the workflow end to end
and obtain compatible results with the author's,
the workflow \may be published.

\subsection{Publishing a workflow}

It is not sufficient to link to a GitHub repository from a journal article.
GitHub provides no guarantees of long-term stability of repositories,
or even of the addresses to them.
Instead,
similarly to data releases,
we \must publish workflows using a service that provides a persistent identifier,
and a commitment to long-term availability.

At time of writing,
the TELOS Collaboration publishes its workflows using Zenodo~\cite{zenodo}.

After a release has been published,
the repository \may still be used for ongoing work.
For example,
if a referee requests changes requiring modifications to the analysis workflow,
these \must be made in the same repository.
These changes \must be released on Zenodo,
as a new version of the same record as above,
before the paper is published.
If work continues on a direct followup paper,
the new paper's analysis \should be a direct continuation of the original's,
in the same repository,
and \should be published as a new version of the same record.

\subsubsection{Preparing an archive}

In principle,
one could connect GitHub with Zenodo,
and automatically generate a record from a GitHub Release.
However,
as discussed above,
this does not include the contents of submodules,
meaning that the archives downloaded from the release would not be usable.

To ensure that all submodules are correctly included in the release,
we \should clone a fresh copy of the repository,
and archive it,
keeping only the working copy,
i.e. excluding the \filename{.git} directories:

\begin{verbatim}
mkdir tmp
cd tmp
git clone --recurse-submodules git@github.com:telos-collaboration/workflow_name
zip -9 --exclude "**/.git/*" --exclude "**/.git"  -r workflow_name workflow_name
\end{verbatim}

One \may instead work from the existing working copy,
but in this case they \must first strip out all data not in the repository,
for example using \verb|git clean|.
Data from the data release \mustnot be included in the workflow release.

\subsubsection{What to include in a release}

The release \should include two files:

\begin{itemize}
  \item
        The \readme file from the repository,
        which \should also form the basis of the Zenodo description, and
  \item
        The ZIP archive of the repository.
\end{itemize}

\subsubsection{Completing the upload form}

The procedure for this is largely the same as that used for data releases,
described in Sec.~\ref{sec:publish-data},
with the following exceptions:

\begin{itemize}
  \item
        The Resource Type \should be ``Workflow''.
  \item
        The Title \should be the title of the corresponding article,
        followed by ``---Analysis workflow''.
  \item
        The License \should match the one agreed and specified in the repository.
  \item
        In addition to the preprint and published version,
        the Related works \must refer to the data release,
        with relation ``requires''.
  \item
        The Software section \must include
        \begin{itemize}
          \item
                A link to the GitHub repository under the \verb|telos-collaboration| organisation.
          \item
                A listing of the programming languages used,
                which are likely to minimally include
                \program{Python} and \program{Snakemake}.
          \item
                The Development Status ``Inactive''.
        \end{itemize}
\end{itemize}

\section{HPC Software}\label{sec:hpc-software}

This area has to date received less focus than
the topics of the preceding sections of this document;
as such,
many elements described as ``\should'' are aspirational
rather than describing current practice.

\subsection{Open Source and Community Software}

The TELOS collaboration has benefited greatly from open source and community software for HPC\@.
We have made extensive use of HiRep~\cite{hirep,DelDebbio:2008zf},
Grid~\cite{grid,Yamaguchi:2022feu},
and Hadrons~\cite{hadrons,antonin_portelli_2023_8023716}.
It is incumbent on us to not only be passive consumers of this software,
but also to give back in support of its continued development and sustainability.

All publications making use of a piece of software
\must specify what software was used,
and what version.
Where we have made modifications to a piece of software for our use case,
these modifications \must be publicly available.
In the absence of this,
others will not be able to reproduce our work,
and the software's authors will not be able to correctly judge
the reach and impact of their work.
In principle,
there \should be a Zenodo release of the specific version used,
to ensure that there is a persistent identifier and long term retention;
however,
to date this has not been done due to
a lack of clarity around appropriate authorship for the resulting Zenodo record.

After making a change to a piece of software,
the developer \should think carefully about whether
it is valuable for the change to  be fed back to the upstream repository.
In instances where the change is likely to be an obstruction to others' use cases,
this is likely not a good idea,
but if the change is neutral or positive,
then minimally,
the change \should be proposed to the upstream maintainer.
If the original author is amenable,
then a pull request
(or equivalent)
\should be submitted,
and appropriate work done in collaboration with the maintainer
to bring the work to a mergeable state.
(The change \may also be proposed directly in the form of a pull request.)

Where changes need to be maintained separately,
we \must retain these changes in a local fork of the repository.
The number of different local forks of a repository \should be minimised;
where possible,
changes \should be merged into a single main branch,
and where this is impractical,
versions \should be retained as branches of a single repository.

\subsection{Data formats}

In general,
software running on HPC \should output data in a format easily ingested by downstream tooling,
and easily shared in a FAIR way.
This \may be in addition to text-based logs;
the logs \should however not be the only data output format.

More specifically,
\begin{itemize}
  \item
        Software generating configurations
        \should output field configurations in the ILDG binary format~\cite{ildg-binary},
        and corresponding metadata in the QCDml~\cite{Maynard:2004wg}\footnote{
        An update to these to support
        higher representations and gauge groups beyond $\mathrm{SU}(3)$
        is in preparation at time of writing,
        but is anticipated to become available very shortly.
        }.
        At time of writing,
        in the versions in use within the TELOS Collaboration,
        neither Grid nor HiRep is able to do this;
        HiRep cannot write the ILDG format at all\footnote{
          The HiRep binary format is compatible with
          the configuration data portion of the ILDG binary format
          in some instances.
        },
        instead using its own format exclusively,
        while Grid can write it but does not set relevant metadata correctly.
  \item
        Software performing observable computations on configurations
        \should output the results in HDF5, JSON, or XML format,
        with sufficient and appropriate metadata to identify the ensemble and computation performed.
        At time of writing,
        Hadrons is able to output in HDF5 and XML format,
        but the specific metadata output are not sufficient for our use case.
        Grid can in principle output these formats,
        but the tools we make use of do not do this.
        HiRep has support for JSON in certain tools in certain forks,
        but our production version does not support this.
\end{itemize}

\subsection{Workflows}

Historically,
a typical HPC workflow has involved setting up some or all of:

\begin{itemize}
  \item
        Writing or adjusting job scripts and input files individually for each ensemble,
  \item
        Manually changing input files after successive jobs,
  \item
        Handcrafting \program{awk} scripts at the command line to check progress and acceptance,
  \item
        Regularly resubmitting jobs as time limits on running jobs elapse,
  \item
        Writing fragile shell scripts to automate one or more aspects of the above,
  \item
        Needing to cancel and resubmit jobs due to an error introduced by one of the above.
\end{itemize}

None of these is desirable;
they are all a drain on time that could have been spent more productively,
they are all barriers to reproducibility,
and they all have the potential to partially or totally invalidate claimed results.
For example,
neglecting to switch the \verb|rlx_start| flag from \verb|new| to \verb|continue|
in \program{HiRep}
introduces sizeable autocorrelations when running the RHMC algorithm.

To avoid these issues,
we \should have two additional tools available to us:
a HMC launcher/monitor
(tentatively called \program{hmcdj})
and a job coordinator
(\program{PlateSpinner}).
At time of writing,
neither of these is written;
we discuss below some requirements to inform their development.

\subsubsection{\program{hmcdj}}

The anticipated responsibilities of this software are:

\begin{itemize}
  \item
        Read parameters specifying the ensemble from a file in a standard format
        (for example, JSON).
  \item
        Identify whether configurations already exist matching these parameters,
        and resume from this if so.
        Otherwise,
        start from a cold, warm, or hot configuration as specified.
  \item
        Be able to start a new chain by resuming from a checkpoint
        but re-seeding the random number generator.
  \item
        Be able to track acceptance of the algorithm over time,
        adjust this automatically during a thermalisation period,
        and raise warnings and refuse to continue if it remains unacceptably low
        or if it changes outside of the thermalisation period.
  \item
        Save provenance and metadata with each configuration,
        as discussed above.
        This \should make it clear,
        for example,
        at what point a secondary chain branched off its parent,
        and when parameters such as the trajectory length were adjusted.
  \item
        Seed the generator deterministically from the ensemble parameters,
        while still providing sufficient entropy to maintain statistical independence of ensembles.
  \item
        Monitor the time taken to complete each trajectory,
        read the remaining time in a job from the environment,
        and if there will be insufficient time to complete the next trajectory,
        checkpoint and quit cleanly instead.
  \item
        Based on this,
        be able to provide an estimate of the notice needed from the scheduler to cleanly quit,
        and take advantage of shorter allocations by allowing preemption with sufficient notice.
        Listen for interrupt signals from the scheduler,
        and checkpoint and cleanly exit after the end of the current trajectory in this case.
\end{itemize}

Grid's HMC implementation satisfies some but not all of these requirements.
We anticipate that \program{hmcdj} could be written as a thin wrapper around Grid's HMC class.

\subsubsection{\program{PlateSpinner}}

A difficulty with some HPC systems is that
HMC is not parallel in Monte Carlo time,
and so one job in a sequence cannot start before the previous one completes.
To avoid jobs starting and then needing to immediately abort,
this ultimately means that successive jobs cannot begin queuing until
their predecessors complete.
This means there can be significant downtime between successive jobs in a sequence,
making it more difficult to obtain the desired statistics in a given time.

However,
where many ensembles are being generated concurrently,
at the point where a single HPC job starts,
it is likely that one of the ensembles is not being actively generated,
or that there is analysis work on one or more ensembles outstanding.
To this end,
we anticipate \program{PlateSpinner} will in its most basic form:

\begin{itemize}
  \item
        Be launched at the start of each job on the machine.
  \item
        Read a database of target generation workloads,
        from the \program{hmcdj} input files.
  \item
        Identify which of these workloads are already running.
  \item
        If there is a workload not already running,
        which is compatible with the job's resources
        (i.e. not so large as to run out of memory,
        or so small as to give very poor parallelisation),
        start it.
  \item
        If all generation workloads are already running,
        identify analysis workloads that are outstanding,
        and not already running,
        and start one of these.
  \item
        Repeat the previous step until
        the anticipated runtime exceeds the remaining time for the job.
\end{itemize}

In addition to the above,
our work utilises machines with GPU resources.
Most such machines also have a non-negligible number of CPU cores available.
Typically,
the majority of these cores are not utilised by the GPU workloads.
Some of our analysis workloads are not able to efficiently utilise GPUs,
so require CPU resources.
Tests show that these can be run concurrently on the same nodes as our GPU workloads
without affecting the performance of the latter.
This gives significantly better utilisation of the resources,
both in terms of resource time allocation
and in terms of total energy utilisation.
We anticipate that \program{PlateSpinner} will also manage this process.

The requirements for \program{PlateSpinner} are similar to the capabilities of \program{Taxi}~\cite{Ayyar:2018wwf};
however,
the latter appears more tightly coupled to HMC generation with
the specific tooling used by the authors of Ref.~\cite{Ayyar:2018wwf}.

\section{Acknowledgements}

The author is grateful for the support of all members of the TELOS Collaboration
in formulating and adopting these guidelines.

This work has been funded by the
UKRI Science and Technologies Facilities Council (STFC)
Research Software Engineering Fellowship EP/V052489/1,
by STFC under Consolidated Grant No. ST/X000648/1,
by the ExaTEPP project EP/X017168/1,
and by
the project to form a Collaborative Computational Project
in Theoretical and Experimental Particle Physics
(CCP-TEPP).

\paragraph*{Open access statement}
For the purpose of open access,
the authors have
applied a Creative Commons Attribution (CC BY) licence
to any Author Accepted Manuscript version arising.

\paragraph*{Data Availability Statement}
No new data were produced in preparation or in support of this work.

\paragraph*{Software Availability Statement}
No new software was produced in preparation or in support of this work.

\nocite{apsrev41Control}
\bibliography{references,revtex-custom}

\begin{thebibliography}{49}%
\makeatletter
\providecommand \@ifxundefined [1]{%
 \@ifx{#1\undefined}
}%
\providecommand \@ifnum [1]{%
 \ifnum #1\expandafter \@firstoftwo
 \else \expandafter \@secondoftwo
 \fi
}%
\providecommand \@ifx [1]{%
 \ifx #1\expandafter \@firstoftwo
 \else \expandafter \@secondoftwo
 \fi
}%
\providecommand \natexlab [1]{#1}%
\providecommand \enquote  [1]{``#1''}%
\providecommand \bibnamefont  [1]{#1}%
\providecommand \bibfnamefont [1]{#1}%
\providecommand \citenamefont [1]{#1}%
\providecommand \href@noop [0]{\@secondoftwo}%
\providecommand \href [0]{\begingroup \@sanitize@url \@href}%
\providecommand \@href[1]{\@@startlink{#1}\@@href}%
\providecommand \@@href[1]{\endgroup#1\@@endlink}%
\providecommand \@sanitize@url [0]{\catcode `\\12\catcode `\$12\catcode
  `\&12\catcode `\#12\catcode `\^12\catcode `\_12\catcode `\%12\relax}%
\providecommand \@@startlink[1]{}%
\providecommand \@@endlink[0]{}%
\providecommand \url  [0]{\begingroup\@sanitize@url \@url }%
\providecommand \@url [1]{\endgroup\@href {#1}{\urlprefix }}%
\providecommand \urlprefix  [0]{URL }%
\providecommand \Eprint [0]{\href }%
\providecommand \doibase [0]{http://dx.doi.org/}%
\providecommand \selectlanguage [0]{\@gobble}%
\providecommand \bibinfo  [0]{\@secondoftwo}%
\providecommand \bibfield  [0]{\@secondoftwo}%
\providecommand \translation [1]{[#1]}%
\providecommand \BibitemOpen [0]{}%
\providecommand \bibitemStop [0]{}%
\providecommand \bibitemNoStop [0]{.\EOS\space}%
\providecommand \EOS [0]{\spacefactor3000\relax}%
\providecommand \BibitemShut  [1]{\csname bibitem#1\endcsname}%
\let\auto@bib@innerbib\@empty
%</preamble>
\bibitem [{tel()}]{telos}%
  \BibitemOpen
  \href@noop {} {\enquote {\bibinfo {title} {{TELOS Collaboration}},}\
  }\bibinfo {howpublished}
  {\url{https://telos-collaboration.github.io}}\BibitemShut {NoStop}%
\bibitem [{\citenamefont {Bennett}(2025{\natexlab{a}})}]{this-github}%
  \BibitemOpen
  \bibfield  {author} {\bibinfo {author} {\bibfnamefont {E.}~\bibnamefont
  {Bennett}},\ }\href@noop {} {\enquote {\bibinfo {title} {The {TELOS
  Collaboration} approach to reproducibility and open science},}\ }\bibinfo
  {howpublished} {\url{https://github.com/telos-collaboration/strategy}}
  (\bibinfo {year} {2025}{\natexlab{a}})\BibitemShut {NoStop}%
\bibitem [{\citenamefont {Bennett}(2025{\natexlab{b}})}]{this-zenodo}%
  \BibitemOpen
  \bibfield  {author} {\bibinfo {author} {\bibfnamefont {E.}~\bibnamefont
  {Bennett}},\ }\href {\doibase 10.5281/zenodo.15113710} {\enquote {\bibinfo
  {title} {The {TELOS Collaboration} approach to reproducibility and open
  science},}\ }\bibinfo {howpublished} {DOI:
  \url{https://doi.org/10.5281/zenodo.15113710}} (\bibinfo {year}
  {2025}{\natexlab{b}})\BibitemShut {NoStop}%
\bibitem [{\citenamefont {{The Turing Way
  Community}}(2023)}]{the_turing_way_community_2023_7625728}%
  \BibitemOpen
  \bibfield  {author} {\bibinfo {author} {\bibnamefont {{The Turing Way
  Community}}},\ }\href {\doibase 10.5281/zenodo.7625728} {\enquote {\bibinfo
  {title} {{The Turing Way: A handbook for reproducible, ethical and
  collaborative research}},}\ }\bibinfo {howpublished}
  {\url{https://book.the-turing-way.org/index.html}} (\bibinfo {year}
  {2023})\BibitemShut {NoStop}%
\bibitem [{\citenamefont {Wilkinson}\ \emph {et~al.}(2016)\citenamefont
  {Wilkinson}, \citenamefont {Dumontier}, \citenamefont {Aalbersberg},
  \citenamefont {Appleton}, \citenamefont {Axton}, \citenamefont {Baak},
  \citenamefont {Blomberg}, \citenamefont {Boiten}, \citenamefont
  {da~Silva~Santos}, \citenamefont {Bourne} \emph
  {et~al.}}]{wilkinson2016fair}%
  \BibitemOpen
  \bibfield  {author} {\bibinfo {author} {\bibfnamefont {M.~D.}\ \bibnamefont
  {Wilkinson}}, \bibinfo {author} {\bibfnamefont {M.}~\bibnamefont
  {Dumontier}}, \bibinfo {author} {\bibfnamefont {I.~J.}\ \bibnamefont
  {Aalbersberg}}, \bibinfo {author} {\bibfnamefont {G.}~\bibnamefont
  {Appleton}}, \bibinfo {author} {\bibfnamefont {M.}~\bibnamefont {Axton}},
  \bibinfo {author} {\bibfnamefont {A.}~\bibnamefont {Baak}}, \bibinfo {author}
  {\bibfnamefont {N.}~\bibnamefont {Blomberg}}, \bibinfo {author}
  {\bibfnamefont {J.-W.}\ \bibnamefont {Boiten}}, \bibinfo {author}
  {\bibfnamefont {L.~B.}\ \bibnamefont {da~Silva~Santos}}, \bibinfo {author}
  {\bibfnamefont {P.~E.}\ \bibnamefont {Bourne}},  \emph {et~al.},\ }\bibfield
  {title} {\enquote {\bibinfo {title} {The {FAIR} guiding principles for
  scientific data management and stewardship},}\ }\href {\doibase
  10.1038/sdata.2016.18} {\bibfield  {journal} {\bibinfo  {journal} {Scientific
  data}\ }\textbf {\bibinfo {volume} {3}},\ \bibinfo {pages} {1--9} (\bibinfo
  {year} {2016})}\BibitemShut {NoStop}%
\bibitem [{\citenamefont {{Creative Commons}}()}]{cc-by}%
  \BibitemOpen
  \bibfield  {author} {\bibinfo {author} {\bibnamefont {{Creative Commons}}},\
  }\href {https://creativecommons.org/licenses/by/4.0/} {\enquote {\bibinfo
  {title} {Attribution 4.0 international deed},}\ }\bibinfo {howpublished}
  {\url{https://creativecommons.org/licenses/by/4.0/}}\BibitemShut {NoStop}%
\bibitem [{\citenamefont {{Massachussets Institute of Technology}}()}]{mit}%
  \BibitemOpen
  \bibfield  {author} {\bibinfo {author} {\bibnamefont {{Massachussets
  Institute of Technology}}},\ }\href {https://opensource.org/license/MIT}
  {\enquote {\bibinfo {title} {{The MIT License}},}\ }\bibinfo {howpublished}
  {\url{https://opensource.org/license/MIT}}\BibitemShut {NoStop}%
\bibitem [{\citenamefont {{Free Software Foundation}}()}]{gpl}%
  \BibitemOpen
  \bibfield  {author} {\bibinfo {author} {\bibnamefont {{Free Software
  Foundation}}},\ }\href {https://www.gnu.org/licenses/gpl-3.0.html} {\enquote
  {\bibinfo {title} {{GNU General Public License}},}\ }\bibinfo {howpublished}
  {\url{https://www.gnu.org/licenses/gpl-3.0.html}}\BibitemShut {NoStop}%
\bibitem [{\citenamefont {Bradner}(1997)}]{rfc2119}%
  \BibitemOpen
  \bibfield  {author} {\bibinfo {author} {\bibfnamefont {S.~O.}\ \bibnamefont
  {Bradner}},\ }\href {\doibase 10.17487/RFC2119} {\enquote {\bibinfo {title}
  {{Key words for use in RFCs to Indicate Requirement Levels}},}\ }\bibinfo
  {howpublished} {RFC 2119} (\bibinfo {year} {1997})\BibitemShut {NoStop}%
\bibitem [{\citenamefont {Ginsparg}(2021)}]{ginsparg2021lessons}%
  \BibitemOpen
  \bibfield  {author} {\bibinfo {author} {\bibfnamefont {P.}~\bibnamefont
  {Ginsparg}},\ }\bibfield  {title} {\enquote {\bibinfo {title} {Lessons from
  {arXiv}'s 30 years of information sharing},}\ }\href {\doibase
  10.1038/s42254-021-00360-z} {\bibfield  {journal} {\bibinfo  {journal}
  {Nature Reviews Physics}\ }\textbf {\bibinfo {volume} {3}},\ \bibinfo {pages}
  {602--603} (\bibinfo {year} {2021})}\BibitemShut {NoStop}%
\bibitem [{\citenamefont {{INSPIRE-HEP}}()}]{inspire-authorxml}%
  \BibitemOpen
  \bibfield  {author} {\bibinfo {author} {\bibnamefont {{INSPIRE-HEP}}},\
  }\href@noop {} {\enquote {\bibinfo {title} {{INSPIRE} collaboration author
  lists},}\ }\bibinfo {howpublished}
  {\url{https://github.com/inspirehep/author.xml\#inspire-collaboration-author-lists}}\BibitemShut
  {NoStop}%
\bibitem [{\citenamefont {{TELOS Collaboration}}(2025)}]{resources}%
  \BibitemOpen
  \bibfield  {author} {\bibinfo {author} {\bibnamefont {{TELOS
  Collaboration}}},\ }\href@noop {} {\enquote {\bibinfo {title} {Resources
  repository},}\ }\bibinfo {howpublished}
  {\url{https://github.com/telos-collaboration/resources}} (\bibinfo {year}
  {2025})\BibitemShut {NoStop}%
\bibitem [{\citenamefont {{The HDF Group}}()}]{hdf5}%
  \BibitemOpen
  \bibfield  {author} {\bibinfo {author} {\bibnamefont {{The HDF Group}}},\
  }\href {https://github.com/HDFGroup/hdf5} {\enquote {\bibinfo {title}
  {{Hierarchical Data Format, version 5}},}\ }\BibitemShut {NoStop}%
\bibitem [{\citenamefont {Bennett}\ \emph
  {et~al.}(2024{\natexlab{a}})\citenamefont {Bennett}, \citenamefont {Hong},
  \citenamefont {Hsiao}, \citenamefont {Lee}, \citenamefont {Lin},
  \citenamefont {Lucini}, \citenamefont {Piai},\ and\ \citenamefont
  {Vadacchino}}]{bennett_2024_13819562}%
  \BibitemOpen
  \bibfield  {author} {\bibinfo {author} {\bibfnamefont {E.}~\bibnamefont
  {Bennett}}, \bibinfo {author} {\bibfnamefont {D.~K.}\ \bibnamefont {Hong}},
  \bibinfo {author} {\bibfnamefont {H.}~\bibnamefont {Hsiao}}, \bibinfo
  {author} {\bibfnamefont {J.-W.}\ \bibnamefont {Lee}}, \bibinfo {author}
  {\bibfnamefont {C.-J.~D.}\ \bibnamefont {Lin}}, \bibinfo {author}
  {\bibfnamefont {B.}~\bibnamefont {Lucini}}, \bibinfo {author} {\bibfnamefont
  {M.}~\bibnamefont {Piai}}, \ and\ \bibinfo {author} {\bibfnamefont
  {D.}~\bibnamefont {Vadacchino}},\ }\href {\doibase 10.5281/zenodo.13819562}
  {\enquote {\bibinfo {title} {Meson spectroscopy in the {Sp(4)} gauge theory
  with three antisymmetric fermions---data release},}\ } (\bibinfo {year}
  {2024}{\natexlab{a}})\BibitemShut {NoStop}%
\bibitem [{\citenamefont {Cone}()}]{markdown-guide}%
  \BibitemOpen
  \bibfield  {author} {\bibinfo {author} {\bibfnamefont {M.}~\bibnamefont
  {Cone}},\ }\href {https://www.markdownguide.org} {\enquote {\bibinfo {title}
  {Markdown guide},}\ }\bibinfo {howpublished}
  {\url{https://www.markdownguide.org}}\BibitemShut {NoStop}%
\bibitem [{\citenamefont {{GitHub}}()}]{github-markdown}%
  \BibitemOpen
  \bibfield  {author} {\bibinfo {author} {\bibnamefont {{GitHub}}},\ }\href
  {https://docs.github.com/en/get-started/writing-on-github/working-with-advanced-formatting/writing-mathematical-expressions}
  {\enquote {\bibinfo {title} {Writing mathematical expressions},}\ }\bibinfo
  {howpublished}
  {\url{https://docs.github.com/en/get-started/writing-on-github/working-with-advanced-formatting/writing-mathematical-expressions}}\BibitemShut
  {NoStop}%
\bibitem [{\citenamefont {InvenioRDM}()}]{zenodo}%
  \BibitemOpen
  \bibfield  {author} {\bibinfo {author} {\bibnamefont {InvenioRDM}},\
  }\href@noop {} {\enquote {\bibinfo {title} {Zenodo},}\ }\bibinfo
  {howpublished} {\url{https://www.zenodo.org}}\BibitemShut {NoStop}%
\bibitem [{\citenamefont {Poelen}\ \emph {et~al.}()\citenamefont {Poelen},
  \citenamefont {Casero}, \citenamefont {Ghaleb}, \citenamefont {Wang},\ and\
  \citenamefont {Welborn}}]{zenodo-upload}%
  \BibitemOpen
  \bibfield  {author} {\bibinfo {author} {\bibfnamefont {J.}~\bibnamefont
  {Poelen}}, \bibinfo {author} {\bibfnamefont {R.}~\bibnamefont {Casero}},
  \bibinfo {author} {\bibfnamefont {T.}~\bibnamefont {Ghaleb}}, \bibinfo
  {author} {\bibfnamefont {S.}~\bibnamefont {Wang}}, \ and\ \bibinfo {author}
  {\bibfnamefont {S.}~\bibnamefont {Welborn}},\ }\href
  {https://github.com/jhpoelen/zenodo-upload} {\enquote {\bibinfo {title}
  {zenodo-upload},}\ }\BibitemShut {NoStop}%
\bibitem [{\citenamefont {{International Lattice Data
  Grid}}()}]{ildg-organization}%
  \BibitemOpen
  \bibfield  {author} {\bibinfo {author} {\bibnamefont {{International Lattice
  Data Grid}}},\ }\href@noop {} {\enquote {\bibinfo {title} {Organization of
  {ILDG} activities},}\ }\bibinfo {howpublished}
  {\url{https://hpc.desy.de/ildg/organization/}},\ \bibinfo {note} {accessed
  2024-08-06}\BibitemShut {NoStop}%
\bibitem [{\citenamefont {Munk}\ \emph {et~al.}(2019)\citenamefont {Munk},
  \citenamefont {Koziar}, \citenamefont {Leinweber}, \citenamefont {Silva}
  \emph {et~al.}}]{swc-git}%
  \BibitemOpen
  \bibfield  {author} {\bibinfo {author} {\bibfnamefont {M.}~\bibnamefont
  {Munk}}, \bibinfo {author} {\bibfnamefont {K.}~\bibnamefont {Koziar}},
  \bibinfo {author} {\bibfnamefont {K.}~\bibnamefont {Leinweber}}, \bibinfo
  {author} {\bibfnamefont {R.}~\bibnamefont {Silva}},  \emph {et~al.},\ }\href
  {\doibase 10.5281/zenodo.3264950} {\enquote {\bibinfo {title}
  {{swcarpentry/git-novice: Software Carpentry: Version Control with Git, June
  2019}},}\ }\bibinfo {howpublished}
  {\url{https://swcarpentry.github.io/git-novice}} (\bibinfo {year}
  {2019})\BibitemShut {NoStop}%
\bibitem [{\citenamefont {Chacon}\ and\ \citenamefont
  {Straub}(2014)}]{chacon2014pro}%
  \BibitemOpen
  \bibfield  {author} {\bibinfo {author} {\bibfnamefont {S.}~\bibnamefont
  {Chacon}}\ and\ \bibinfo {author} {\bibfnamefont {B.}~\bibnamefont
  {Straub}},\ }\href {https://git-scm.com/book} {\enquote {\bibinfo {title}
  {Pro {Git}},}\ } (\bibinfo {year} {2014})\BibitemShut {NoStop}%
\bibitem [{\citenamefont {Druskat}(2023)}]{cff}%
  \BibitemOpen
  \bibfield  {author} {\bibinfo {author} {\bibfnamefont {S.}~\bibnamefont
  {Druskat}},\ }\href@noop {} {\enquote {\bibinfo {title} {{Citation File
  Format}},}\ }\bibinfo {howpublished}
  {\url{https://citation-file-format.github.io/}} (\bibinfo {year}
  {2023})\BibitemShut {NoStop}%
\bibitem [{\citenamefont {Druskat}()}]{cff-init}%
  \BibitemOpen
  \bibfield  {author} {\bibinfo {author} {\bibfnamefont {S.}~\bibnamefont
  {Druskat}},\ }\href@noop {} {\enquote {\bibinfo {title} {{cffinit}},}\
  }\bibinfo {howpublished}
  {\url{https://citation-file-format.github.io/cff-initializer-javascript/}}\BibitemShut
  {NoStop}%
\bibitem [{\citenamefont {GitHub}()}]{gitignore-python}%
  \BibitemOpen
  \bibfield  {author} {\bibinfo {author} {\bibnamefont {GitHub}},\ }\href@noop
  {} {\enquote {\bibinfo {title} {Python .gitignore template},}\ }\bibinfo
  {howpublished}
  {\url{https://github.com/github/gitignore/blob/main/Python.gitignore}}\BibitemShut
  {NoStop}%
\bibitem [{\citenamefont {pre commit}()}]{pre-commit}%
  \BibitemOpen
  \bibfield  {author} {\bibinfo {author} {\bibnamefont {pre commit}},\
  }\href@noop {} {\enquote {\bibinfo {title} {pre-commit},}\ }\bibinfo
  {howpublished} {\url{https://pre-commit.com}}\BibitemShut {NoStop}%
\bibitem [{\citenamefont {pre-commit CI}()}]{pre-commit-ci}%
  \BibitemOpen
  \bibfield  {author} {\bibinfo {author} {\bibnamefont {pre-commit CI}},\
  }\href@noop {} {\enquote {\bibinfo {title} {pre-commit ci},}\ }\bibinfo
  {howpublished} {\url{https://pre-commit.ci}}\BibitemShut {NoStop}%
\bibitem [{\citenamefont {M{\"o}lder}\ \emph {et~al.}(2021)\citenamefont
  {M{\"o}lder} \emph {et~al.}}]{molder2021sustainable}%
  \BibitemOpen
  \bibfield  {author} {\bibinfo {author} {\bibfnamefont {F.}~\bibnamefont
  {M{\"o}lder}} \emph {et~al.},\ }\bibfield  {title} {\enquote {\bibinfo
  {title} {Sustainable data analysis with snakemake},}\ }\href@noop {}
  {\bibfield  {journal} {\bibinfo  {journal} {F1000Research}\ }\textbf
  {\bibinfo {volume} {10}} (\bibinfo {year} {2021})}\BibitemShut {NoStop}%
\bibitem [{\citenamefont {Collins}\ \emph {et~al.}()\citenamefont {Collins}
  \emph {et~al.}}]{carpentries-snakemake}%
  \BibitemOpen
  \bibfield  {author} {\bibinfo {author} {\bibfnamefont {D.}~\bibnamefont
  {Collins}} \emph {et~al.},\ }\href@noop {} {\enquote {\bibinfo {title} {{Tame
  Your Workflow with Snakemake}},}\ }\bibinfo {howpublished}
  {\url{https://carpentries-incubator.github.io/workflows-snakemake}}\BibitemShut
  {NoStop}%
\bibitem [{\citenamefont {Koester}\ \emph {et~al.}()\citenamefont {Koester}
  \emph {et~al.}}]{snakemake-tutorial}%
  \BibitemOpen
  \bibfield  {author} {\bibinfo {author} {\bibfnamefont {J.}~\bibnamefont
  {Koester}} \emph {et~al.},\ }\href@noop {} {\enquote {\bibinfo {title}
  {Tutorial: General use},}\ }\bibinfo {howpublished}
  {\url{https://snakemake.readthedocs.io/en/stable/tutorial/tutorial.html}}\BibitemShut
  {NoStop}%
\bibitem [{\citenamefont {{TELOS Collaboration}}()}]{workflow-template}%
  \BibitemOpen
  \bibfield  {author} {\bibinfo {author} {\bibnamefont {{TELOS
  Collaboration}}},\ }\href@noop {} {\enquote {\bibinfo {title} {{TELOS
  Collaboration} analysis workflow template},}\ }\bibinfo {howpublished}
  {\url{https://github.com/telos-collaboration/workflow_template}}\BibitemShut
  {NoStop}%
\bibitem [{\citenamefont {Bennett}()}]{fme}%
  \BibitemOpen
  \bibfield  {author} {\bibinfo {author} {\bibfnamefont {E.}~\bibnamefont
  {Bennett}},\ }\href@noop {} {\enquote {\bibinfo {title}
  {format\_multiple\_errors},}\ }\bibinfo {howpublished}
  {\url{https://github.com/edbennett/format_multiple_errors}}\BibitemShut
  {NoStop}%
\bibitem [{\citenamefont {Groth}\ and\ \citenamefont {Moreau}(2013)}]{prov}%
  \BibitemOpen
  \bibfield  {author} {\bibinfo {author} {\bibfnamefont {P.}~\bibnamefont
  {Groth}}\ and\ \bibinfo {author} {\bibfnamefont {L.}~\bibnamefont {Moreau}},\
  }\href@noop {} {\enquote {\bibinfo {title} {{PROV-Overview}},}\ }\bibinfo
  {howpublished} {\url{https://www.w3.org/TR/prov-overview/}} (\bibinfo {year}
  {2013})\BibitemShut {NoStop}%
\bibitem [{\citenamefont {Auge}\ \emph {et~al.}(2023)\citenamefont {Auge},
  \citenamefont {Bali}, \citenamefont {Klettke}, \citenamefont {Lud\"ascher},
  \citenamefont {S\"oldner}, \citenamefont {Weish\"aupl},\ and\ \citenamefont
  {Wettig}}]{Auge:2023vnd}%
  \BibitemOpen
  \bibfield  {author} {\bibinfo {author} {\bibfnamefont {T.}~\bibnamefont
  {Auge}}, \bibinfo {author} {\bibfnamefont {G.}~\bibnamefont {Bali}}, \bibinfo
  {author} {\bibfnamefont {M.}~\bibnamefont {Klettke}}, \bibinfo {author}
  {\bibfnamefont {B.}~\bibnamefont {Lud\"ascher}}, \bibinfo {author}
  {\bibfnamefont {W.}~\bibnamefont {S\"oldner}}, \bibinfo {author}
  {\bibfnamefont {S.}~\bibnamefont {Weish\"aupl}}, \ and\ \bibinfo {author}
  {\bibfnamefont {T.}~\bibnamefont {Wettig}},\ }\bibfield  {title} {\enquote
  {\bibinfo {title} {{Provenance for Lattice QCD workflows}},}\ \ }(\bibinfo
  {year} {2023})\ \Eprint {http://arxiv.org/abs/2303.12640} {arXiv:2303.12640
  [hep-lat]} \BibitemShut {NoStop}%
\bibitem [{\citenamefont {Bennett}\ \emph
  {et~al.}(2024{\natexlab{b}})\citenamefont {Bennett}, \citenamefont {Hong},
  \citenamefont {Hsiao}, \citenamefont {Lee}, \citenamefont {Lin},
  \citenamefont {Lucini}, \citenamefont {Piai},\ and\ \citenamefont
  {Vadacchino}}]{bennett_2024_13819431}%
  \BibitemOpen
  \bibfield  {author} {\bibinfo {author} {\bibfnamefont {E.}~\bibnamefont
  {Bennett}}, \bibinfo {author} {\bibfnamefont {D.~K.}\ \bibnamefont {Hong}},
  \bibinfo {author} {\bibfnamefont {H.}~\bibnamefont {Hsiao}}, \bibinfo
  {author} {\bibfnamefont {J.-W.}\ \bibnamefont {Lee}}, \bibinfo {author}
  {\bibfnamefont {C.-J.~D.}\ \bibnamefont {Lin}}, \bibinfo {author}
  {\bibfnamefont {B.}~\bibnamefont {Lucini}}, \bibinfo {author} {\bibfnamefont
  {M.}~\bibnamefont {Piai}}, \ and\ \bibinfo {author} {\bibfnamefont
  {D.}~\bibnamefont {Vadacchino}},\ }\href {\doibase 10.5281/zenodo.13819431}
  {\enquote {\bibinfo {title} {Meson spectroscopy in the sp(4) gauge theory
  with three antisymmetric fermions---analysis workflow},}\ } (\bibinfo {year}
  {2024}{\natexlab{b}})\BibitemShut {NoStop}%
\bibitem [{\citenamefont {Lepage}\ and\ \citenamefont {Gohlke}()}]{lsqfit}%
  \BibitemOpen
  \bibfield  {author} {\bibinfo {author} {\bibfnamefont {P.}~\bibnamefont
  {Lepage}}\ and\ \bibinfo {author} {\bibfnamefont {C.}~\bibnamefont
  {Gohlke}},\ }\href@noop {} {\enquote {\bibinfo {title} {lsqfit},}\ }\bibinfo
  {howpublished} {\url{https://github.com/gplepage/lsqfit}}\BibitemShut
  {NoStop}%
\bibitem [{\citenamefont {Lepage}\ and\ \citenamefont
  {Gohlke}(2024)}]{peter_lepage_2024_12690493}%
  \BibitemOpen
  \bibfield  {author} {\bibinfo {author} {\bibfnamefont {P.}~\bibnamefont
  {Lepage}}\ and\ \bibinfo {author} {\bibfnamefont {C.}~\bibnamefont
  {Gohlke}},\ }\href {\doibase 10.5281/zenodo.12690493} {\enquote {\bibinfo
  {title} {gplepage/lsqfit: lsqfit version 13.2.3},}\ }\bibinfo {howpublished}
  {\href{https://doi.org/10.5281/zenodo.12690493}{doi:10.5281/zenodo.12690493}}
  (\bibinfo {year} {2024})\BibitemShut {NoStop}%
\bibitem [{\citenamefont {Lepage}()}]{corrfitter}%
  \BibitemOpen
  \bibfield  {author} {\bibinfo {author} {\bibfnamefont {P.}~\bibnamefont
  {Lepage}},\ }\href@noop {} {\enquote {\bibinfo {title} {corrfitter},}\
  }\bibinfo {howpublished}
  {\url{https://github.com/gplepage/corrfitter}}\BibitemShut {NoStop}%
\bibitem [{\citenamefont {Lepage}(2021)}]{peter_lepage_2021_5733391}%
  \BibitemOpen
  \bibfield  {author} {\bibinfo {author} {\bibfnamefont {P.}~\bibnamefont
  {Lepage}},\ }\href {\doibase 10.5281/zenodo.5733391} {\enquote {\bibinfo
  {title} {gplepage/corrfitter: corrfitter version 8.2},}\ }\bibinfo
  {howpublished}
  {\href{https://doi.org/10.5281/zenodo.5733391}{doi:10.5281/zenodo.5733391}}
  (\bibinfo {year} {2021})\BibitemShut {NoStop}%
\bibitem [{\citenamefont {{DVC}}()}]{dvc}%
  \BibitemOpen
  \bibfield  {author} {\bibinfo {author} {\bibnamefont {{DVC}}},\ }\href@noop
  {} {\enquote {\bibinfo {title} {{Data Version Control}},}\ }\bibinfo
  {howpublished} {\url{https://dvc.org}}\BibitemShut {NoStop}%
\bibitem [{\citenamefont {Pica}\ \emph {et~al.}()\citenamefont {Pica} \emph
  {et~al.}}]{hirep}%
  \BibitemOpen
  \bibfield  {author} {\bibinfo {author} {\bibfnamefont {C.}~\bibnamefont
  {Pica}} \emph {et~al.},\ }\href@noop {} {\enquote {\bibinfo {title}
  {{HiRep}},}\ }\bibinfo {howpublished}
  {\url{https://github.com/claudiopica/HiRep}}\BibitemShut {NoStop}%
\bibitem [{\citenamefont {Del~Debbio}\ \emph {et~al.}(2010)\citenamefont
  {Del~Debbio}, \citenamefont {Patella},\ and\ \citenamefont
  {Pica}}]{DelDebbio:2008zf}%
  \BibitemOpen
  \bibfield  {author} {\bibinfo {author} {\bibfnamefont {L.}~\bibnamefont
  {Del~Debbio}}, \bibinfo {author} {\bibfnamefont {A.}~\bibnamefont {Patella}},
  \ and\ \bibinfo {author} {\bibfnamefont {C.}~\bibnamefont {Pica}},\
  }\bibfield  {title} {\enquote {\bibinfo {title} {{Higher representations on
  the lattice: Numerical simulations. SU(2) with adjoint fermions}},}\ }\href
  {\doibase 10.1103/PhysRevD.81.094503} {\bibfield  {journal} {\bibinfo
  {journal} {Phys. Rev. D}\ }\textbf {\bibinfo {volume} {81}},\ \bibinfo
  {pages} {094503} (\bibinfo {year} {2010})},\ \Eprint
  {http://arxiv.org/abs/0805.2058} {arXiv:0805.2058 [hep-lat]} \BibitemShut
  {NoStop}%
\bibitem [{\citenamefont {Boyle}\ \emph {et~al.}()\citenamefont {Boyle} \emph
  {et~al.}}]{grid}%
  \BibitemOpen
  \bibfield  {author} {\bibinfo {author} {\bibfnamefont {P.}~\bibnamefont
  {Boyle}} \emph {et~al.},\ }\href@noop {} {\enquote {\bibinfo {title}
  {{Grid}},}\ }\bibinfo {howpublished}
  {\url{https://github.com/paboyle/Grid}}\BibitemShut {NoStop}%
\bibitem [{\citenamefont {Yamaguchi}\ \emph {et~al.}(2022)\citenamefont
  {Yamaguchi}, \citenamefont {Boyle}, \citenamefont {Cossu}, \citenamefont
  {Filaci}, \citenamefont {Lehner},\ and\ \citenamefont
  {Portelli}}]{Yamaguchi:2022feu}%
  \BibitemOpen
  \bibfield  {author} {\bibinfo {author} {\bibfnamefont {A.}~\bibnamefont
  {Yamaguchi}}, \bibinfo {author} {\bibfnamefont {P.}~\bibnamefont {Boyle}},
  \bibinfo {author} {\bibfnamefont {G.}~\bibnamefont {Cossu}}, \bibinfo
  {author} {\bibfnamefont {G.}~\bibnamefont {Filaci}}, \bibinfo {author}
  {\bibfnamefont {C.}~\bibnamefont {Lehner}}, \ and\ \bibinfo {author}
  {\bibfnamefont {A.}~\bibnamefont {Portelli}},\ }\bibfield  {title} {\enquote
  {\bibinfo {title} {{Grid: OneCode and FourAPIs}},}\ }\href {\doibase
  10.22323/1.396.0035} {\bibfield  {journal} {\bibinfo  {journal} {PoS}\
  }\textbf {\bibinfo {volume} {LATTICE2021}},\ \bibinfo {pages} {035} (\bibinfo
  {year} {2022})},\ \Eprint {http://arxiv.org/abs/2203.06777} {arXiv:2203.06777
  [hep-lat]} \BibitemShut {NoStop}%
\bibitem [{\citenamefont {Portelli}\ \emph {et~al.}()\citenamefont {Portelli}
  \emph {et~al.}}]{hadrons}%
  \BibitemOpen
  \bibfield  {author} {\bibinfo {author} {\bibfnamefont {A.}~\bibnamefont
  {Portelli}} \emph {et~al.},\ }\href@noop {} {\enquote {\bibinfo {title}
  {{Hadrons}},}\ }\bibinfo {howpublished}
  {\url{https://github.com/aportelli/Hadrons}}\BibitemShut {NoStop}%
\bibitem [{\citenamefont {Portelli}\ \emph {et~al.}(2023)\citenamefont
  {Portelli} \emph {et~al.}}]{antonin_portelli_2023_8023716}%
  \BibitemOpen
  \bibfield  {author} {\bibinfo {author} {\bibfnamefont {A.}~\bibnamefont
  {Portelli}} \emph {et~al.},\ }\href {\doibase 10.5281/zenodo.8023716}
  {\enquote {\bibinfo {title} {aportelli/hadrons: Hadrons v1.4},}\ }\bibinfo
  {howpublished}
  {\href{https://doi.org/10.5281/zenodo.8023716}{doi:10.5281/zenodo.8023716}}
  (\bibinfo {year} {2023})\BibitemShut {NoStop}%
\bibitem [{\citenamefont {{ILDG Metadata Working Group}}(2005)}]{ildg-binary}%
  \BibitemOpen
  \bibfield  {author} {\bibinfo {author} {\bibnamefont {{ILDG Metadata Working
  Group}}},\ }\href@noop {} {\enquote {\bibinfo {title} {{ILDG Binary File
  Format (Rev. 1.1)}},}\ }\bibinfo {howpublished}
  {\url{https://www-zeuthen.desy.de/apewww/ILDG/specifications/ildg-file-format-1.1.pdf}}
  (\bibinfo {year} {2005})\BibitemShut {NoStop}%
\bibitem [{\citenamefont {Maynard}\ and\ \citenamefont
  {Pleiter}(2005)}]{Maynard:2004wg}%
  \BibitemOpen
  \bibfield  {author} {\bibinfo {author} {\bibfnamefont {C.~M.}\ \bibnamefont
  {Maynard}}\ and\ \bibinfo {author} {\bibfnamefont {D.}~\bibnamefont
  {Pleiter}},\ }\bibfield  {title} {\enquote {\bibinfo {title} {{QCDml: First
  milestone for building an International Lattice Data Grid}},}\ }\href
  {\doibase 10.1016/j.nuclphysbps.2004.11.116} {\bibfield  {journal} {\bibinfo
  {journal} {Nucl. Phys. B Proc. Suppl.}\ }\textbf {\bibinfo {volume} {140}},\
  \bibinfo {pages} {213--221} (\bibinfo {year} {2005})},\ \Eprint
  {http://arxiv.org/abs/hep-lat/0409055} {arXiv:hep-lat/0409055} \BibitemShut
  {NoStop}%
\bibitem [{\citenamefont {Ayyar}\ \emph {et~al.}(2018)\citenamefont {Ayyar},
  \citenamefont {Hackett}, \citenamefont {Jay},\ and\ \citenamefont
  {Neil}}]{Ayyar:2018wwf}%
  \BibitemOpen
  \bibfield  {author} {\bibinfo {author} {\bibfnamefont {V.}~\bibnamefont
  {Ayyar}}, \bibinfo {author} {\bibfnamefont {D.~C.}\ \bibnamefont {Hackett}},
  \bibinfo {author} {\bibfnamefont {W.~I.}\ \bibnamefont {Jay}}, \ and\
  \bibinfo {author} {\bibfnamefont {E.~T.}\ \bibnamefont {Neil}},\ }\bibfield
  {title} {\enquote {\bibinfo {title} {{Automated lattice data generation}},}\
  }\href {\doibase 10.1051/epjconf/201817509009} {\bibfield  {journal}
  {\bibinfo  {journal} {EPJ Web Conf.}\ }\textbf {\bibinfo {volume} {175}},\
  \bibinfo {pages} {09009} (\bibinfo {year} {2018})},\ \Eprint
  {http://arxiv.org/abs/1802.00851} {arXiv:1802.00851 [hep-lat]} \BibitemShut
  {NoStop}%
\bibitem [{\citenamefont {Vanschoren}\ \emph {et~al.}(2013)\citenamefont
  {Vanschoren}, \citenamefont {van Rijn}, \citenamefont {Bischl},\ and\
  \citenamefont {Torgo}}]{OpenML2013}%
  \BibitemOpen
  \bibfield  {author} {\bibinfo {author} {\bibfnamefont {J.}~\bibnamefont
  {Vanschoren}}, \bibinfo {author} {\bibfnamefont {J.~N.}\ \bibnamefont {van
  Rijn}}, \bibinfo {author} {\bibfnamefont {B.}~\bibnamefont {Bischl}}, \ and\
  \bibinfo {author} {\bibfnamefont {L.}~\bibnamefont {Torgo}},\ }\bibfield
  {title} {\enquote {\bibinfo {title} {Openml: networked science in machine
  learning},}\ }\href {\doibase 10.1145/2641190.2641198} {\bibfield  {journal}
  {\bibinfo  {journal} {SIGKDD Explorations}\ }\textbf {\bibinfo {volume}
  {15}},\ \bibinfo {pages} {49--60} (\bibinfo {year} {2013})}\BibitemShut
  {NoStop}%
\end{thebibliography}%

\clearpage
\appendix

\section{Checklist for publishing a journal article}\label{app:checklist}

In this Appendix is a checklist that \should be completed for all publications.
Underneath each high-level goal
(circles)
is a set of tasks that \should be completed before the goal can be completed
(squares).

\begin{CheckList}{Goal}
  \Goal{open}{Circulate draft to collaboration}
  \begin{CheckList}{Task}
    \Task{open}{DOI for data release obtained and added to draft}
    \Task{open}{DOI for analysis release obtained and added to draft}
    \Task{open}{Hand-generated provisional/placeholder plots marked}
    \Task{open}{Software used is committed to version control}
    \Task{open}{Software names and versions (e.g.\ commit IDs) are specified}
    \Task{open}{Rights retention statement is present}
    \Task{open}{Data availability statement is present}
    \Task{open}{Software availability statement is present}
  \end{CheckList}
  \Goal{open}{Publish data release}
  \begin{CheckList}{Task}
    \Task{open}{\readme contains data format descriptions for all files}
    \Task{open}{Raw data are all included}
    \Task{open}{CSVs are present for all data presented in paper (tabulated or plotted)}
    \Task{open}{Archives do not contain unwanted files (e.g.\ operating system metadata files like \filename{.DS\_Store})}
    \Task{open}{Release cross-checked with analysis workflow by another collaboration member}
  \end{CheckList}
  \Goal{open}{Publish analysis workflow}
  \begin{CheckList}{Task}
    \Task{open}{\readme includes requirements installation instructions}
    \Task{open}{\readme includes instructions for getting input data}
    \Task{open}{\readme includes instructions for running workflow}
    \Task{open}{\readme notes expected run time of workflow}
    \Task{open}{
      No data are hardcoded into code.
      (Check for any numbers with more than three significant figures.)
    }
    \Task{open}{Quoted data carry appropriate attribution}
    \Task{open}{Software environment is specified in one or more \filename{.yml} files}
    \Task{open}{Workflow runs end to end from a single command without errors}
    \Task{open}{
      Workflow archive does not include any unwanted files
      (e.g.\ data files, or operating system metadata files like \filename{.DS\_Store})
    }
    \Task{open}{Workflow cross-checked with data release by another collaboration member}
    \Task{open}{GitHub repository is public}
  \end{CheckList}
  \Goal{open}{Publish pre-print on arXiv}
  \begin{CheckList}{Task}
    \Task{open}{\authorxml written, validated against schema, and included in source package}
    \Task{open}{Data release public \emph{(optional)}}
    \Task{open}{Analysis workflow release public \emph{(optional)}}
    \Task{open}{Final analysis workflow is committed to version control}
    \Task{open}{
      All plots, tables, and quoted numbers generated from workflow.
      No placeholder markers remain.
    }
    \Task{open}{All assets regenerated from a clean workflow run}
  \end{CheckList}
  \Goal{open}{Submit manuscript to journal}
  \begin{CheckList}{Task}
    \Task{open}{Preprint added to website}
    \Task{open}{arXiv paper password forwarded to coauthors}
  \end{CheckList}
  \Goal{open}{Submit corrected manuscript to journal after implementing referee feedback}
  \begin{CheckList}{Task}
    \Goal{open}{
      If changes have been made to analysis or presentation,
      all assets regenerated from a clean workflow run.
    }
    \Task{open}{Updated analysis workflow is committed to version control}
  \end{CheckList}
  \Goal{open}{Proofs are approved with journal, article is published}
  \begin{CheckList}{Task}
    \Task{open}{Author Accepted Manuscript uploaded to institutional repository (e.g.\ Cronfa)}
    \Task{open}{Finalised data release public \emph{(required)}}
    \Task{open}{Finalised analysis workflow release public \emph{(required)}}
    \Task{open}{
      If updated data/workflow releases published,
      manuscript updated to point to DOI of updated version.
    }
  \end{CheckList}
  \Goal{open}{Publication process finalised}
  \begin{CheckList}{Task}
    \Task{open}{Website updated with published article}
    \Task{open}{Institutional repository (e.g.\ Cronfa) updated with journal details}
  \end{CheckList}
\end{CheckList}

\clearpage
\section{Applying these techniques outside of lattice}\label{app:nonlattice}

The guidance in this document has been written
with the context of lattice quantum field theory in mind,
as well as the specific requirements and capabilities of the TELOS Collaboration.
For other contexts,
there is a wide range of advice available;
for example,
The Turing Way~\cite{the_turing_way_community_2023_7625728} provides
tools and techniques applicable across much of data science and research software,
while OpenML~\cite{OpenML2013}
provides tooling and resources for open research using machine learning techniques.

That said,
there is a subset of the advice presented here that can be summarised
to form a minimal set of principles applicable,
for example,
in work in theoretical physics not on the lattice.

\subsection{Data}

If a publication generates new data,
these data \must be released openly, citably, and in machine-readable form.
``Data'' in this context includes:
\begin{itemize}
  \item One or more numbers with error bars,
  \item One or more points plotted on a graph,
  \item Numbers needing a data table to display in a paper, and
  \item Coefficients in equations with more than a handful of terms.
\end{itemize}
As a rule of thumb,
if numbers would be useful for others,
and taking it from the text of the paper would be laborious
or result in loss of precision,
then it is data and \must be released.

These numbers \must be shared in CSV format,
unless there is a strong reason to do otherwise.
It is worth thinking carefully about how to combine related data
to minimise the number of separate files that a reader has to work with;
this might differ from the native representations you currently use.
Where a metadata schema is available,
this \must be followed to ensure interoperability and reusability;
regardless,
a common format for data between publications \should be aimed for.
The guidance in Section~\ref{sec:dr-numbers} provides more detail on this.

The data release \must also include documentation
so that the data consumer is able to make use of it.
A plain-text, human-readable
(for example, Markdown)
\readme file \should be designed as the first
(and potentially only)
place a reader looks.
The \readme \must provide at least a link to the paper presenting the data,
and a listing describing the contents of each file in the release.
It \must either describe or link to documentation describing
the data formats of the other files in the release.
Section~\ref{sec:dr-documentation} discusses in more detail how to approach this.

The data release \must be published in a repository that commits to long-term availability
and that offers a persistent identifier,
such as Zenodo~\cite{zenodo},
which by default allows 50GiB and 100 files per dataset.
One \may obtain a DOI from Zenodo to include in a manuscript
in advance of publishing the data;
details on this are in Section~\ref{sec:get-doi}.
Section~\ref{sec:upload-form} provides
a brief guide to completing the Zenodo upload form for lattice data.

\subsection{Workflows}

A first step in the direction of open workflows
is to share the tools needed to translate
from the data shared in the data release
to the outputs in the publication
(plots, tables, etc.,
as discussed in Section~\ref{sec:assets});
this allows a reader to better understand the data formats,
and to at least verify that the data do indeed give the presented results.
If shared in this way,
it likely makes more sense to include this as a part of the data release,
rather than releasing a separate workflow,
as the workflow only makes sense in the context of the data it was written against.

However,
as discussed in Section~\ref{sec:workflows},
to fully ensure reproducibility,
sharing the full software workflow to go from input data to output data is needed.
This is something always worth striving for,
even if sometimes one falls short,
for example,
due to not having permission to publish
one or more tools that were shared privately by a colleague.

When it is possible to share the full workflow,
this \should give bitwise identical results
(aside from differences in metadata around,
for example,
when the run was executed)
when re-run on the same platform with the same parallelisation options,
and \must give compatible results when re-run anywhere.
(See Section~\ref{sec:numerical-repro} for additional detail.)
Before it is published,
it \must be tested end to end from a fresh start;
this \should be done by a co-author who did not perform the original analysis.
More details on how to approach this are in Section~\ref{sec:testing}.

More generally,
those working on workflows \should review the work of others
and have their work reviewed in turn,
as discussed in Section~\ref{sec:code-review},
to enable the spread of good practice.
When planning new work,
it is \recommended to adopt version control and good practices for repository structure
(discussed in Sections~\ref{sec:wf-essentials} and~\ref{sec:repository-structure}),
and to consider utilising a workflow manager
(see Section~\ref{sec:wf-manager}),
to handle the flow of data between different tools,
without needing to reinvent the wheel.

\subsection{Publications}

Aside from mentions of high-performance computing which might not be relevant,
all of the discussion of Section~\ref{sec:publications} applies more widely.
Specifically,
all narrative work \should be released under a Creative Commons license;
rights retention statements \may be used to ensure that this is permitted.
If work is not funded by UKRI,
then details of requirements for open access \must be checked with funders,
to ensure that any rights retention statements are valid.
An \authorxml file \may be used
to assist in providing metadata to indexing services like INSPIRE\@.
Software,
computing resources,
open data,
and data storage resources \must be acknowledged,
in addition to funders and those who have provided useful discussions
Software and Data Availability Statements \must be added after the acknowledgments.

\end{document}